\documentclass[a4paper,fleqn,usenatbib]{mnras}

\usepackage{mathptmx}

\usepackage[T1]{fontenc}
\usepackage{ae,aecompl}

\usepackage{graphicx}	
\usepackage{amsmath}	
\usepackage{amssymb}	
\usepackage{rotating}
\usepackage{xspace}

\newcommand{\kms}{km\,s$^{-1}$\xspace} 
\let\oldAA\AA
\renewcommand{\AA}{\text{\oldAA}\xspace} 

\newcommand{\around}{$\sim$}

\newcommand{\Halpha}{H$\alpha$}

\newcommand{\Hbeta}{H$\beta$}

\newcommand{\HI}{H{\sc i}}

\newcommand{\HII}{H{\sc ii}}

\newcommand{\arcdeg}{\(\stackrel{\:\circ}{\textstyle.\rule{0pt}{0.65ex}}\)}

\usepackage{xspace}
\newcommand{\OIII}{\ensuremath{\text{[O\sc iii]}}\xspace}
\newcommand{\OIIIL}{\ensuremath{\text{[O\sc iii]}\lambda5007}\xspace}
\newcommand{\OIIILL}{\ensuremath{\text{[O\sc iii]}\lambda\lambda4959,5007}\xspace}
\newcommand{\NII}{\ensuremath{\text{[N\sc ii]}}}

\newcommand{\orcid}[2]{\href{http://orcid.org/#2}{#1}}

\title[Emission Line Maps of the SMC]{Emission Line Velocity, Metallicity and Extinction Maps of the Small Magellanic Cloud}

\author[Lah et al.]
{\orcid{Philip Lah}{0000-0001-6841-6553}$^{1}$\thanks{E-mail: pkl9469@nyu.edu}, 
\orcid{Matthew Colless}{0000-0001-9552-8075}$^{2,3}$, 
\orcid{Francesco D'Eugenio}{0000-0003-2388-8172}$^{4,5,6}$,
\orcid{Brent Groves}{0000-0002-9768-0246}$^{3,7}$
and \orcid{Joseph D.~Gelfand}{0000-0003-4679-1058}$^{1}$
\\
$^{1}$Center for Astrophysics and Space Science (CASS), New York University Abu Dhabi, PO Box 129188, Abu Dhabi, UAE \\
$^{2}$Research School of Astronomy and Astrophysics, Australian National University, Canberra, ACT 2611, Australia\\
$^{3}$ARC Centre of Excellence for All Sky Astrophysics in 3 Dimensions (ASTRO 3D), Australia\\
$^{4}$Kavli Institute for Cosmology, University of Cambridge, Madingley Road, Cambridge, CB3 0HA, UK\\
$^{5}$Cavendish Laboratory, University of Cambridge, 19 JJ Thomson Avenue, Cambridge, CB3 0HE, UK\\
$^{6}$INAF -- Osservatorio Astronomico di Brera, via Brera 28, I-20121 Milano, Italy\\
$^{7}$International Centre for Radio Astronomy Research, University of Western Australia, 35 Stirling Hwy, Crawley, WA 6009, Australia\\
}

\date{Accepted XXX. Received YYY; in original form ZZZ}

\pubyear{2022}

\begin{document}
\label{firstpage}
\pagerange{\pageref{firstpage}--\pageref{lastpage}}
\maketitle


\begin{abstract}
  Optical emission lines across the Small Magellanic Cloud (SMC) have been measured from multiple fields using the Australian National University (ANU) 2.3m telescope with the Wide-Field Spectrograph (WiFeS). Interpolated maps of the gas-phase metallicity, extinction, \Halpha\ radial velocity and \Halpha\ velocity dispersion have been made from these measurements. There is a metallicity gradient from the centre to the north of the galaxy of \around $-$0.095\,dex/kpc with a shallower metallicity gradient from the centre to the south of the galaxy of \around $-$0.013\,dex/kpc. There is a extinction gradient of \around $-$0.086\,E(B-V)/kpc from the centre going north and shallower going from the centre to the south of \around $-$0.0089\,E(B-V)/kpc. The SMC eastern arm has lower extinction than the main body. The radial velocity of the gas from the \Halpha\ line and the \HI\ line have been compared across the SMC. In general there is good agreement between the two measurements, though there are a few notable exceptions. Both show a region that has different radial velocity to the bulk motion of the SMC in the southern western corner by at least 16\,\kms. The velocity dispersion from \Halpha\ and \HI\ across the SMC have also been compared, with the \Halpha\ velocity dispersion usually the higher of the two. The eastern arm of the SMC generally has lower velocity dispersion than the SMC's main body. These measurements enable a detailed examination of the SMC, highlighting its nature as a disrupted satellite galaxy. 
\end{abstract}


\begin{keywords}
  ISM: abundance, ISM: \HII\ regions, ISM: kinematics and dynamics, galaxies: ISM, Magellanic Clouds
\end{keywords}



\section{Introduction}

The Small Magellanic Cloud (SMC) is a dwarf irregular, star-forming, low-metallicity galaxy with total stellar mass of $\rm 7.5 \times 10^7 M_{\odot}$ \citep{bekki09}. The SMC is interacting with the Large Magellanic Cloud and the Milky Way, and is a highly disrupted system \citep{murai80}. As the SMC is so close, we can study these effects in detail, something that is not possible for more distant galaxies of similar size. 

The stellar metallicity across the SMC has been measured in detail by many authors including \citet{carrera08}, \citet{parisi10}, \citet{kapakos11}, \citet{haschke12}, \citet{piatti12}, \citet{dobbie14}, \citet{piatti15}, \citet{deb15}, \citet{parisi16}, \citet{narloch21} and \citet{parisi22}. A variety of methods have been used, including spectroscopy of the Ca~II triplet and stellar isochrones. Some authors find no stellar metallicity gradient across the SMC \citep{parisi10,kapakos11,piatti12,haschke12,deb15}; others do find a stellar metallicity gradient, though often it is fairly shallow \citep{carrera08,dobbie14,piatti15,parisi16,choudhury18,choudhury20,narloch21,parisi22}. 

Likewise, many authors have examined the properties of \HII\ regions in the SMC, including \citet{meyssonnier93}, \citet{kurt99}, \citet{reyes99}, \citet{garnett99}, \citet{peimbert00}, \citet{vermeij02b}, \citet{testor03}, \citet{lebouteiller08}, \citet{martinhernandez08}, \citet{vanloon10}, \citet{penaguerrero12}, \citet{reyes15}, \citet{toribiosancipriano17} and \citet{jin23}. The UM/CTIO Magellanic Cloud Emission-line Survey used narrowband filters at \OIII, \Halpha\ and [S{\sc ii}] to quantify properties of the interstellar medium of the SMC \citep{smith99}. Most of these observations have only been on a few \HII\ regions, and they employ a variety of methods to determine the gas-phase metallicity. Here we measure the gas-phase metallicity across the entire SMC using a single method in order to study the global properties of this dwarf galaxy.

Dust extinction has been measured across the SMC by authors including \citet{caplan96}, \citet{hutchings01}, \citet{zaritsky02}, \citet{gordon03}, \citet{dobashi09}, \citet{haschke11}, \citet{mericajones17}, \citet{joshi19}, \citet{gorski20}, \citet{skowron21} and \citet{chen22}. These observations span ultraviolet to infrared wavelengths and have mainly measured the dust extinction using stellar observations, many using the intrinsic luminosity of red clump stars. In this work we create an extinction map of the SMC using the ratio of \Halpha\ and \Hbeta\ line fluxes, that is from gas not stars. (Although \citet{caplan96} measured the extinction from \Halpha\ and \Hbeta\ line fluxes, they did not create a map of the SMC from their data.)

\citet{lecoarer93} used a scanning Fabry-Perot interferometer and photon counting system to measure the \Halpha\ emission line across the SMC. They create flux and radial velocity maps of the SMC, but did measure the velocity dispersion of the hot gas. The neutral atomic hydrogen (\HI) 21~cm emission, radial velocity and velocity dispersion across the SMC has been imaged using combined Parkes telescope and Australia Telescope Compact Array (ATCA) observations \citep{staveleysmith95,staveleysmith97,stanimirovic99,stanimirovic04}. Observations of the stellar radial velocity and velocity dispersion have been made by \citet{harris06}, \citet{evans08}, \citet{parisi09}, \citet{depropris10} and \citet{dobbie14b}. Each targeted a different stellar population. They find that the stars in the SMC are in a non-rotating spheroid, in contrast to the \HI\ gas which is in a rotating disc or bar-like structure \citep{stanimirovic04}. While measurements of the radial velocity across the SMC have been made via many methods, no comparison has been made of the hot, \Halpha-emitting gas with the colder \HI\ gas. In addition, the velocity dispersion of the \Halpha\ emitting gas has not been looked at in detail, especially in comparison to the \HI\ gas. \citet{smart19} observed the SMC with the Wisconsin H-alpha Mapper (WHAM) which has a beam size of 1\,degree. They probed the diffuse ionised gas halo of the SMC while we, with our better resolution, are focused on \HII\ regions in the central region of the SMC.   

In this paper we present the first large uniform emission-line survey across the SMC using optical spectroscopy. Section~\ref{Observations} gives the details of the observations. Section~\ref{Gas_Phase_Metallicity} presents the measurements of the gas-phase metallicity from optical emission lines across the SMC. Section~\ref{Extinction} provides a map of the extinction in the SMC from the flux ratio of the \Halpha\ and \Hbeta\ lines. The radial velocity from the \Halpha\ emission line across the SMC is presented in Section~\ref{Radial_Velocity}, and compared to the radial velocity of the \HI\ gas. Similarly, Section~\ref{Velocity_Dispersion} presents the velocity dispersion across the SMC as measured from the \Halpha\ emission line, and compared to the \HI\ velocity dispersion in the same regions. Section~\ref{Conclusion} summarises the results of our work and presents our conclusions. This study is timely and relevant in providing a complement and comparison to the survey being done by the Local Volume Mapper \citep{konidaris20}. 


\section{Observations}
\label{Observations}


\begin{figure*}
  \includegraphics[width=\columnwidth]{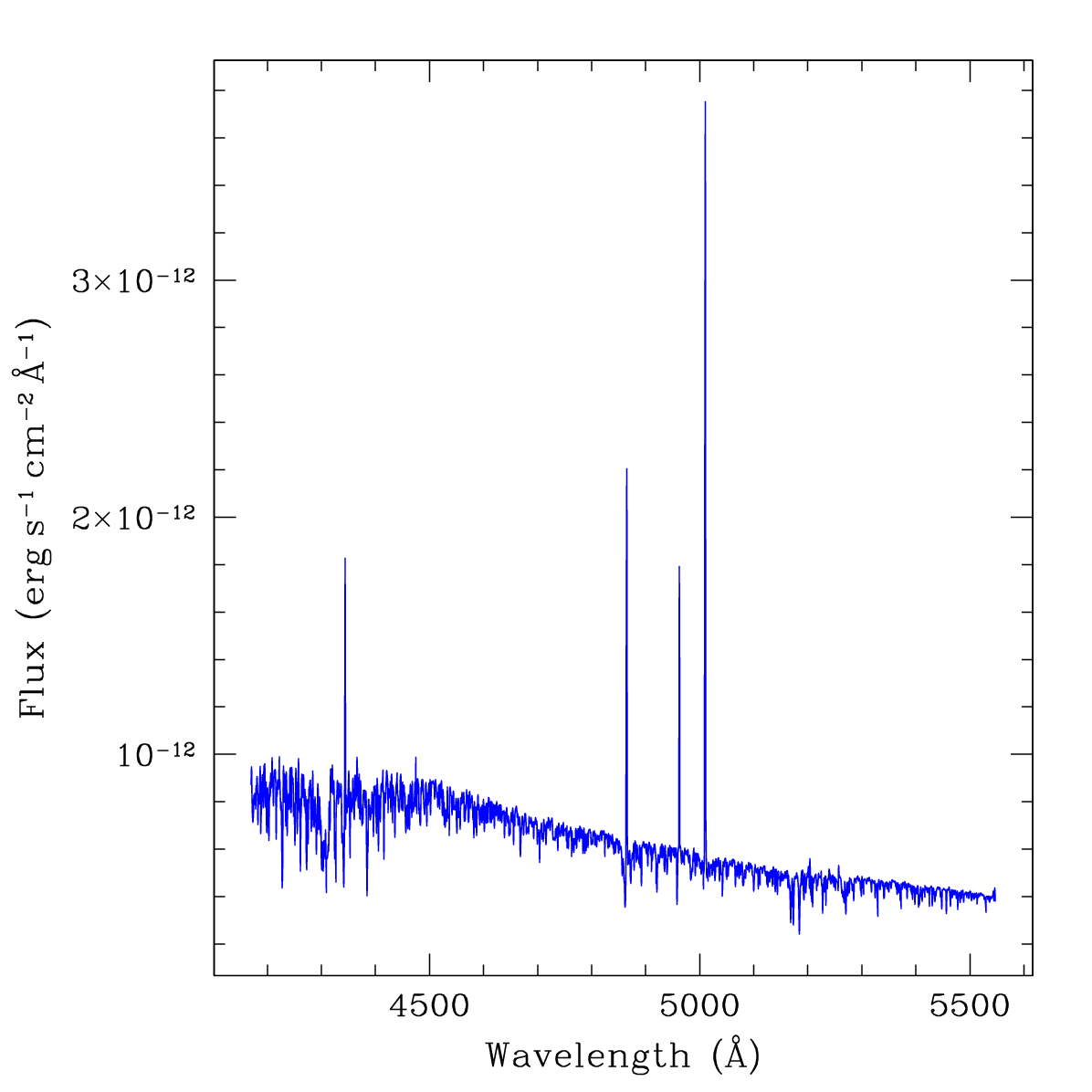}
  \includegraphics[width=\columnwidth]{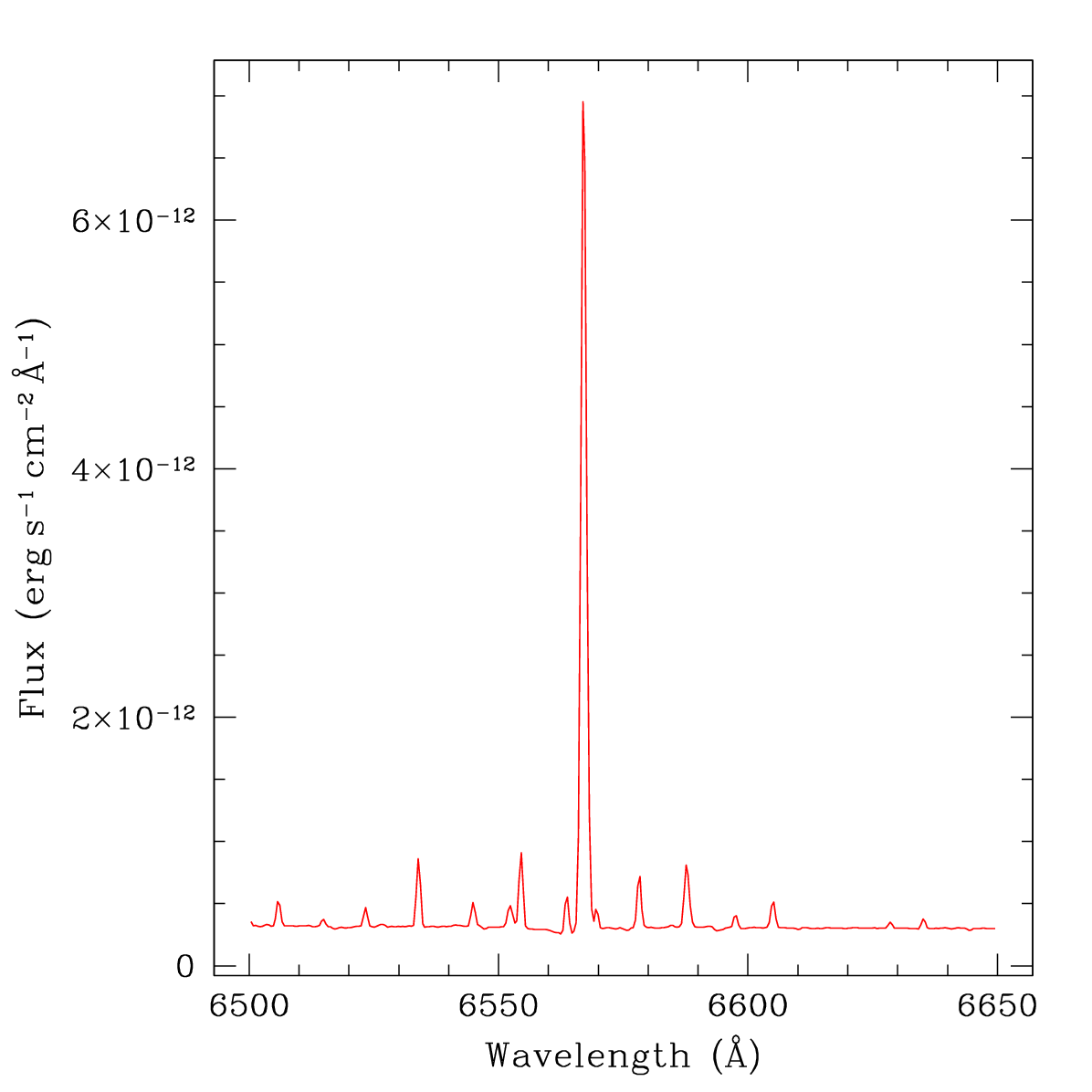}
  \caption{Left: an example blue spectrum for a WiFeS field; the continuum is dominated by moonlight. Right: an example red spectrum for a WiFeS field, in the wavelength region of the \Halpha\ line; the full red spectrum is not shown, as it is dominated by skylines. These spectra come from a field located at R.A.=00:54:18.2 and Dec.=$-$72:48:22 (R.A.=13\arcdeg576 and Dec.=$-$72\arcdeg806).}
  \label{fig:plot_spectrum}
\end{figure*}


\begin{figure*}
  \includegraphics[width=15cm]{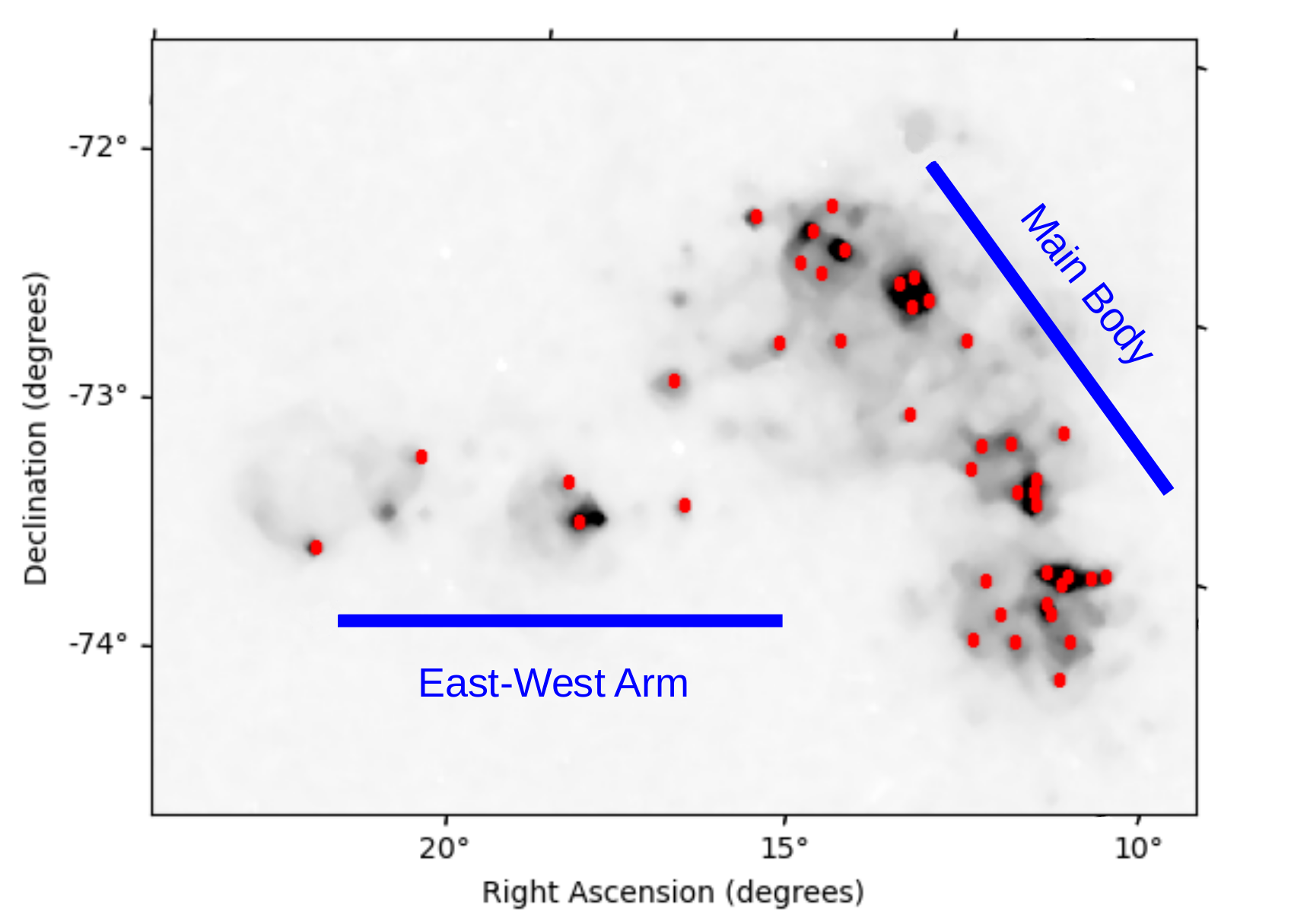}
  \caption{The positions of the 41 WiFeS fields used for this analysis, overlaid on the SHASSA H$\alpha$-emission image for the SMC. The field locations are shown by red points; the coverage of the WiFeS IFU (37\,arcsec $\times$ 25\,arcsec) is considerably smaller than these points.  The east-west arm is the beginning of the Magellanic Bridge that links the SMC to the Large Magellanic Cloud.}
  \label{fig:SMC_vel_region}
\end{figure*}



\begin{figure}
  \includegraphics[width=\columnwidth]{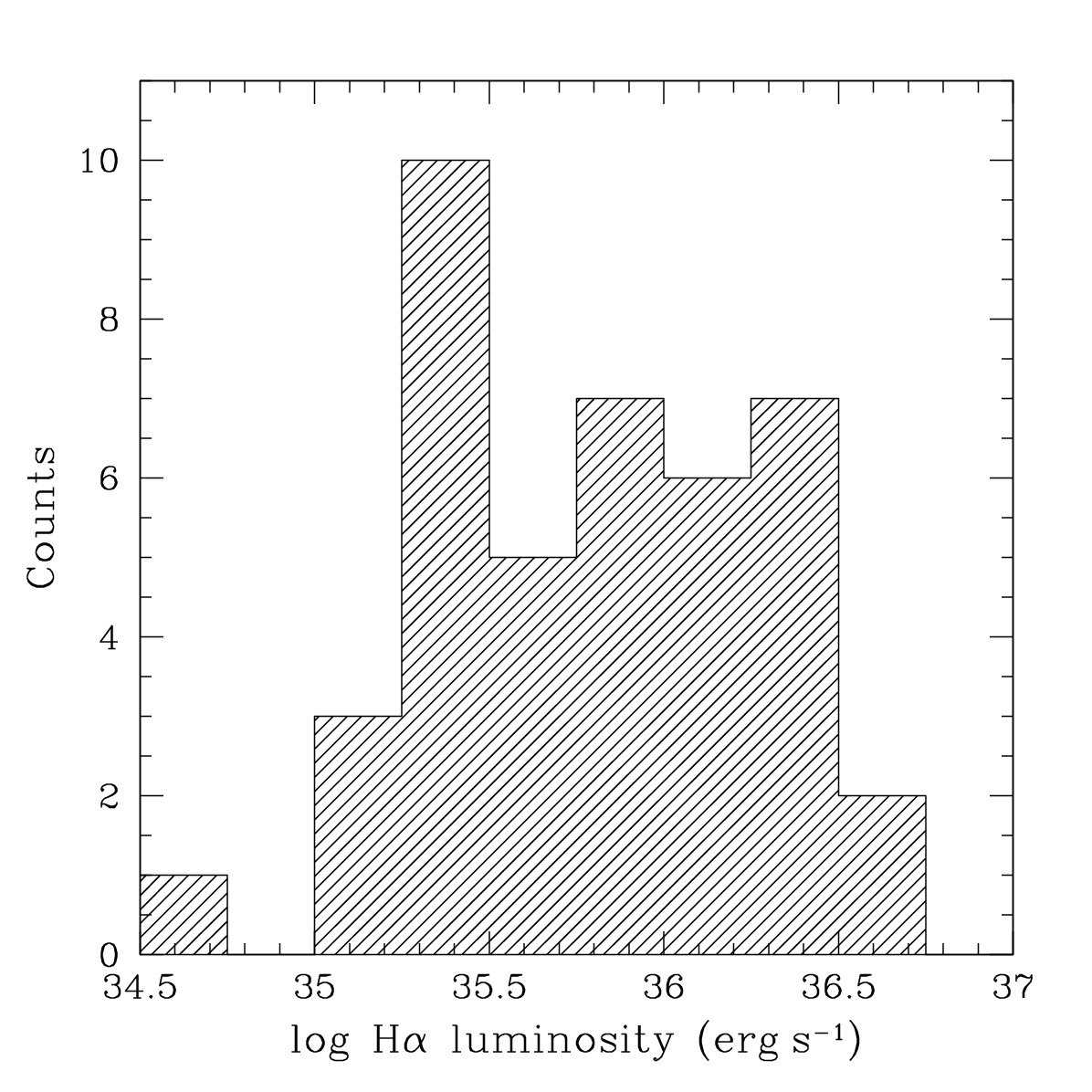}
  \caption{The histogram of the \Halpha\ luminosity for the 41 WiFeS fields with significant \Halpha\ emission measurements. The distance to the SMC is taken to be 62\,kpc \citep{degrijs15}.}
  \label{fig:hist_emission_lines_luminosity_Halpha}
\end{figure}


This paper follows on from our study of the optical emission lines in the Large Magellanic Cloud (LMC) \citep{lah24}. The project used bright time on the ANU 2.3m telescope with the WiFeS spectrograph \citep{dopita07,dopita10}. WiFeS is an integral field spectrographs with field of view 38~$\times$~25~arcsec with spaxel size 1~arcsec (950 spaxels in total). It has both a blue and red spectrographic arms. The idea was to observe multiple fields spread across the SMC and to sum all the pixels in each IFU observation to create a one-dimensional spectrum for each field with high signal-to-noise. 

The WiFeS B7000 ($R = 7000$) and R7000 ($R = 7000$) spectrographic gratings were used.  WiFeS uses a dichroic, so the blue- and red-arm spectra are obtained simultaneously.  The fields to observe were chosen from the Southern H-Alpha Sky Survey Atlas \citep[SHASSA;][]{gaustad01}. SHASSA combines narrow- and broad-band filter observations centred on the \Halpha\ emission line so that the H$\alpha$ emission flux can be measured. The sensitivity level of SHASSA is $1.2 \times 10^{-17}$~erg~s$^{-1}$~cm$^{-2}$~arcsec$^{-2}$. Fields were chosen to give a good coverage of the major \Halpha\ emission regions in the SMC. The integration times were based on the flux values reported by SHASSA, and on the empirical instrument sensitivity we measured on our LMC observations \citep{lah24}. Fields with SHASSA flux greater than $8.5 \times 10^{-15}$~erg~s$^{-1}$~cm$^{-2}$~arcsec$^{-2}$ were observed for 20 minutes; there were 27 such fields. Fields with SHASSA flux between 6 and $8.5 \times 10^{-15}$~erg~s$^{-1}$~cm$^{-2}$~arcsec$^{-2}$ were observed for 40 minutes; there were 18 such fields. This gives a total of 45 fields.  Observations were taken over 6 nights during the southern hemisphere spring of 2021.

The data was reduced with {\sc PyWiFeS} \citep{childress14} using the observed biases, flats, and arcs to give a blue and a red datacube for each of the 45 fields.  The data cubes have wavelength coverage from 4170 to 5548~\AA\ (the blue spectrum) and wavelength coverage from 5400 to 7000~\AA\ (the red spectrum).  At this point the data cubes were summed to give one-dimensional spectra. Simple summing was sufficient, as there was not a large variation in the pixel-to-pixel noise level. The bottom two rows of pixels in the IFU data are bad pixels with a high noise level and were removed. This leaves the IFU summing over 37~$\times$~25~arcsec.  No sky subtraction was done as we were looking at the emission lines above the continuum level. Using the Manual and Automatic Redshifting software \citep[{\sc MARZ},][]{hinton16}, a preliminary radial velocity for each one-dimensional spectrum was determined from the blue spectra. The radial velocities were primarily determined from the H$\beta$\ emission line with confirming lines from the \OIII\ lines at 4959~\AA\ and 5007~\AA. Of the 45 fields, 43 fields had measurable preliminary radial velocities. 

The properties of the emission lines of \Hbeta, \OIIILL, \Halpha\ and \NII$\lambda\lambda6548,6584$ emission lines were measured from the WiFeS spectra using a Python code as described in \citet{lah24}. The two gratings, blue and red, are modelled independently in this code. The formal noise vector is scaled up before the fit to take into account correlations between the pixels. This is estimated from a linear fit to the continuum $\pm 120\text{--}150$~\AA\ either side of the emission line and masking the emission. A robust measure of the standard deviation of the residuals is then calculated. The formal uncertainties are scaled so that their new median in the fitted region is equal to the standard deviation of the residuals. The data, with the scaled noise vector, are then modelled as the sum of a linear continuum and Gaussian emission lines. The lines are integrated over each spectral pixel. Within the measurements \Hbeta, \OIII, \Halpha\ and \NII\ may each have different redshifts and line widths. The two doublets, \OIII\ and \NII, have fixed line ratios as prescribed by atomic physics. For each grating, we therefore have eight free parameters: two for the linear continuum, three for the recombination line, and three for the collisionally excited doublet. In order to reach convergence it was necessary to set the \Hbeta\ redshift to that of \OIII, a line that was usually stronger in the data. The best-fit parameters are found using a simple linear regression algorithm. Thus we have a measurement of the flux of the lines, the radial velocity of the lines, and the full width half maximum (FWHM) of the lines. Bootstrapping the data one hundred times yielded uncertainties on these quantities.

In addition, measurements were made of the bright skylines at 6300.304~\AA\ and 6863.955~\AA. There was an offset in the radial velocity of the measured from the \Halpha\ emission line.  These skyline measurements were used to correct this offset and also to quantify the instrumental resolution. 

Of the 43 fields, 41 had significant \Halpha\ emission measurements. To be significant, the \Halpha\ line had to have a flux in the line at least 3 times the noise in the line, a redshift between 100\,\kms\ and 300\,\kms\ (the rough limits of the SMC radial velocity range) and \Halpha\ Gaussian FWHM $>0.5$\AA\ and $<3$\AA\ (to ensure the line was not a noise spike or a match to the continuum). In addition, the instrumental resolution measured from the skylines at 6300.304~\AA\ and at 6863.955~\AA\ had to have Gaussian FWHM $<3$\AA.

The position of the \Halpha\ emission line fields can be seen in Figure~\ref{fig:SMC_vel_region} overlaid on a SHASSA image of the \HII\ regions of the SMC. The main body of the SMC, where the vast majority of the stars reside, runs north to south in this image. The east-west arm of the SMC can be seen in \Halpha\ emission but contains very few stars. The regions of significant \HI\ emission from observations by \citet{staveleysmith95,staveleysmith97,stanimirovic99,stanimirovic04} run along the main body of the SMC but also extend along the east-west arm, similar to the \HII\ regions.

The \Halpha\ fluxes for the fields, as measured from  their integrated spectra, can be seen in the histogram in Figure~\ref{fig:hist_emission_lines_luminosity_Halpha}. The \Halpha\ fluxes and the field positions are listed in Table~\ref{tab:Halpha} in the Appendix.


\section{Results}
\label{Results}

\subsection{Gas-Phase Metallicity}
\label{Gas_Phase_Metallicity}


\begin{figure}
  \includegraphics[width=\columnwidth]{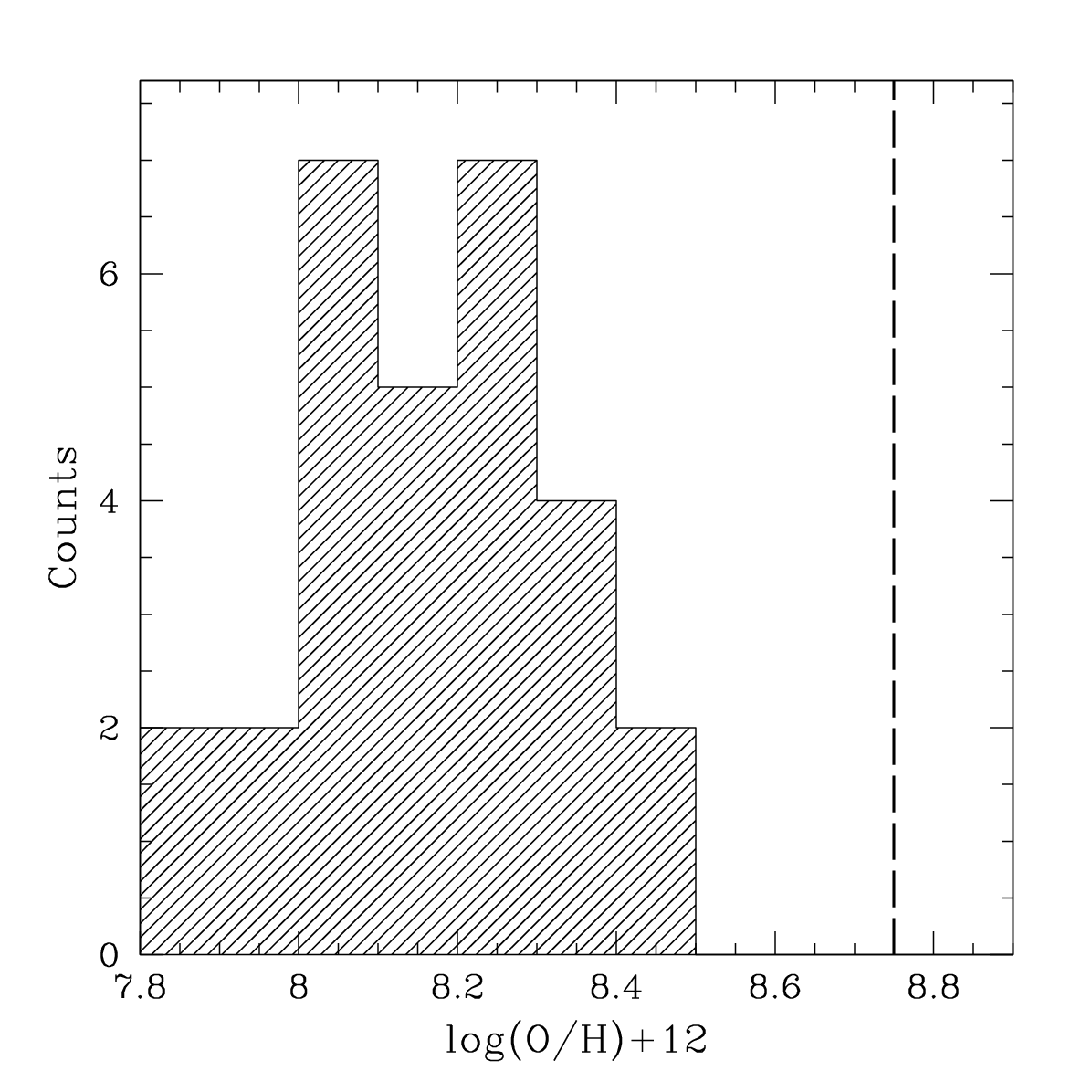}
  \caption{The histogram of the gas-phase metallicity for the 26 WiFeS fields with significant metallicity measurements.  The dashed line at 8.75 is the solar oxygen abundance from \citet{bergemann21}.}
  
  \label{fig:hist_emission_lines_new_Kewely}
\end{figure}


\begin{figure*}
  \centerline{\includegraphics[height=7cm]{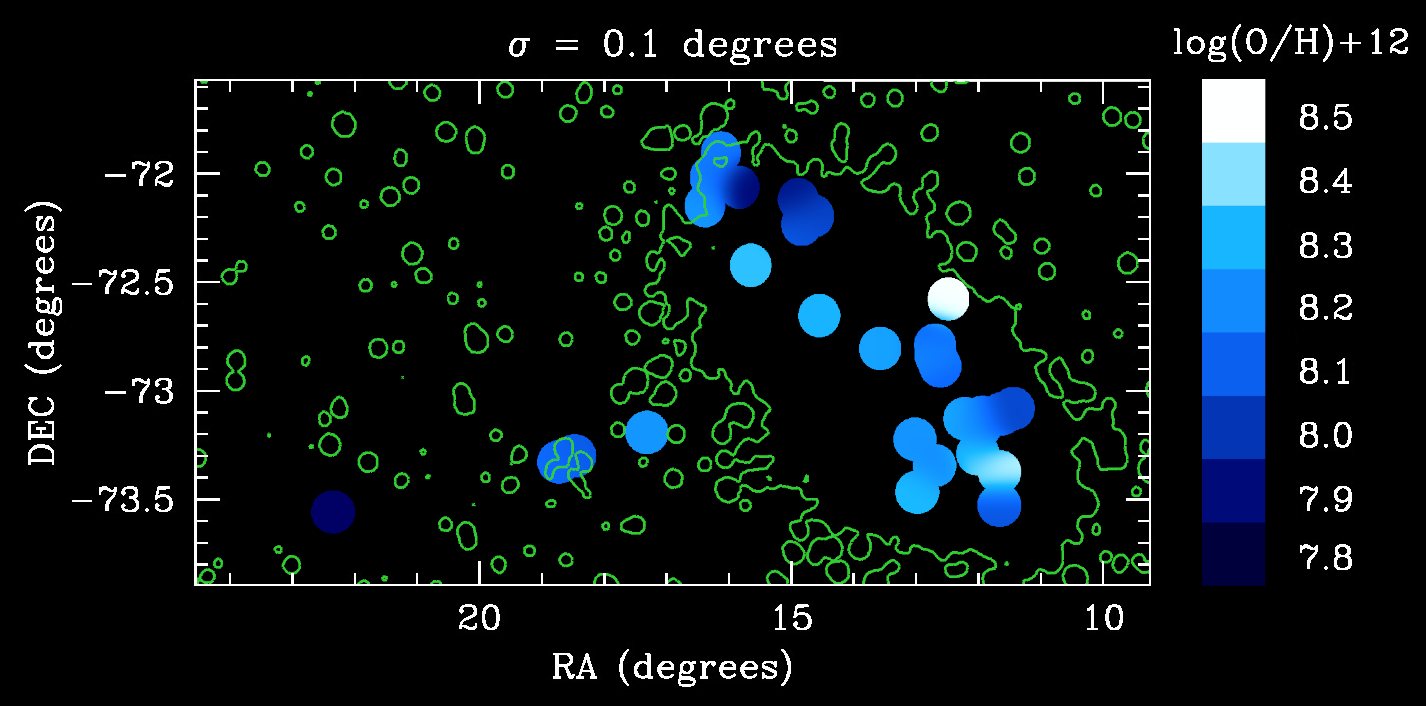}}
  \centerline{\includegraphics[height=7cm]{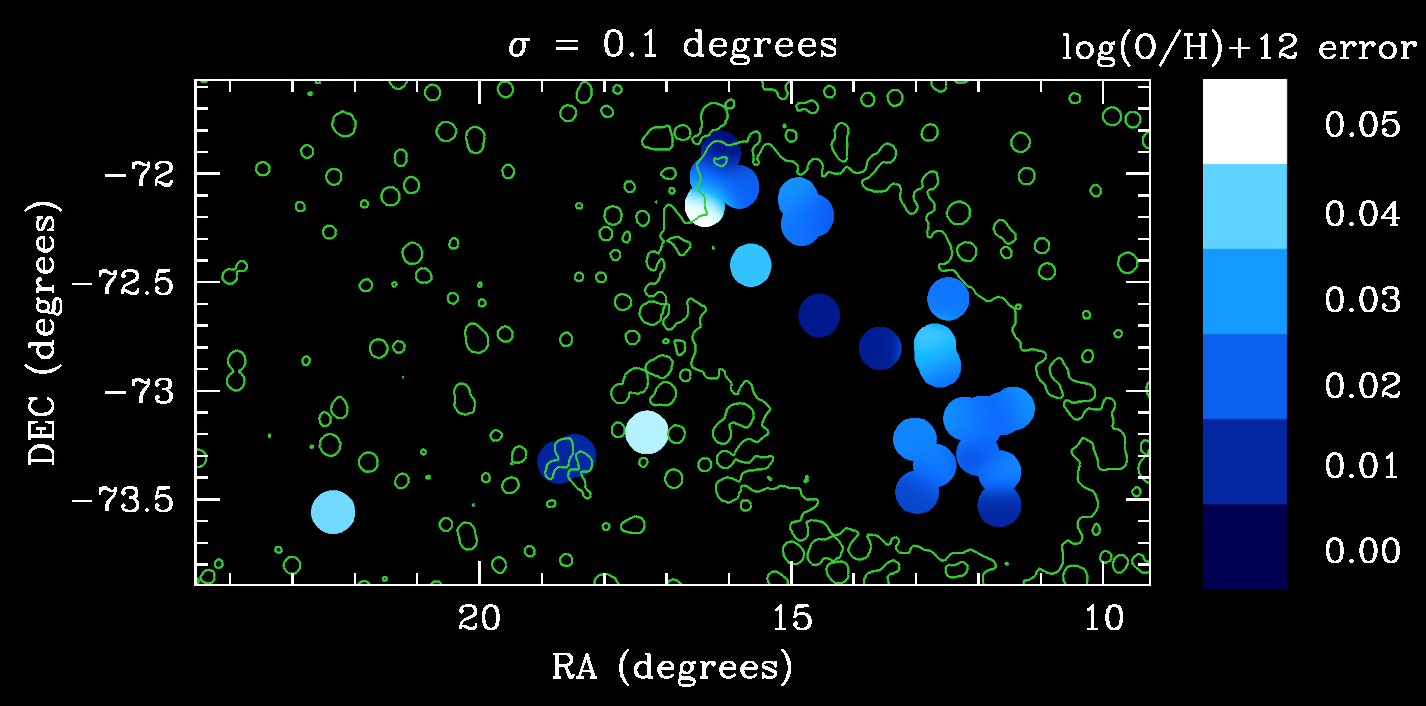}}
  \centerline{\includegraphics[height=7cm]{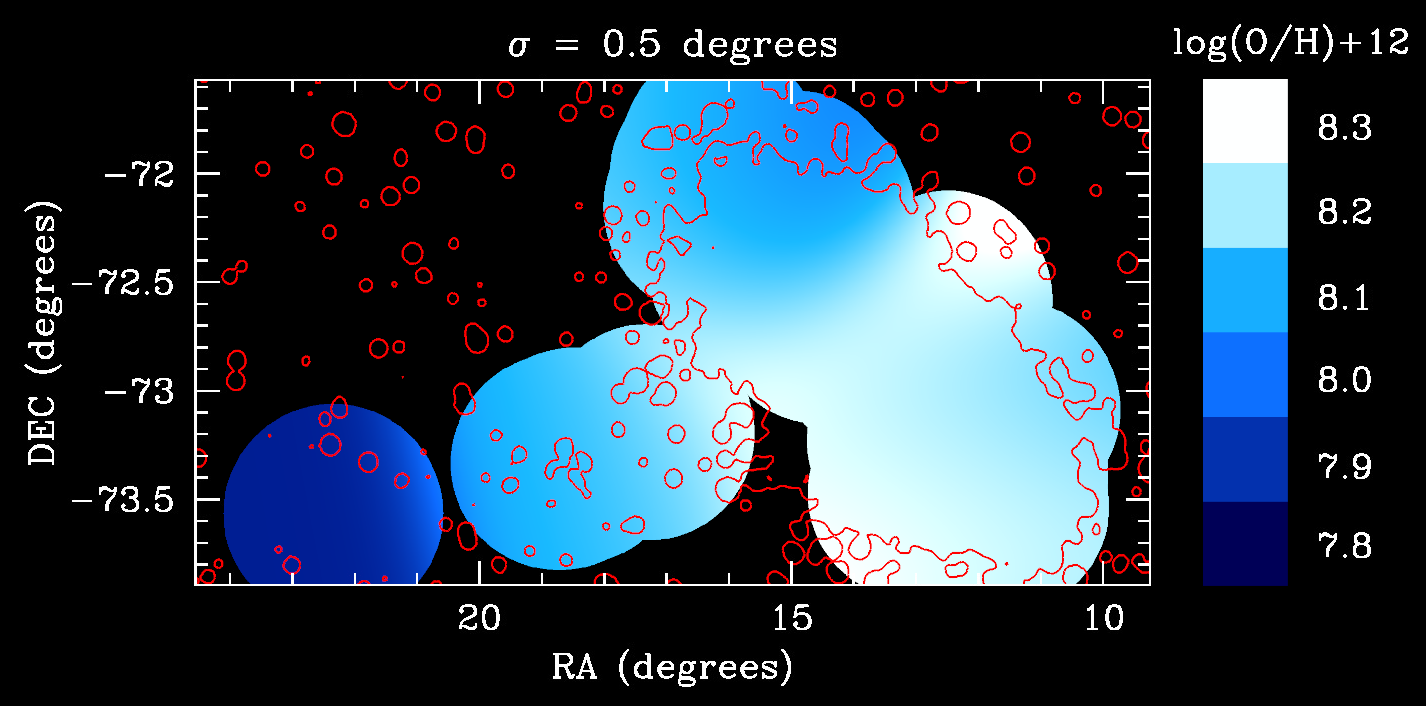}}
  \caption{The SMC gas-phase metallicity interpolated with a Gaussian of size 0.1~degrees (top panel) and 0.5~degrees (bottom panel). The middle panel shows the error distribution for the metallicity measurements. In each case the cutoff radius is equal to the Gaussian smoothing size.  The contours are from the SHASSA R band and are used to highlight the location of the SMC main body.}
  \label{fig:metallicity}
\end{figure*}


\begin{figure*}
  \centerline{\includegraphics[height=7cm]{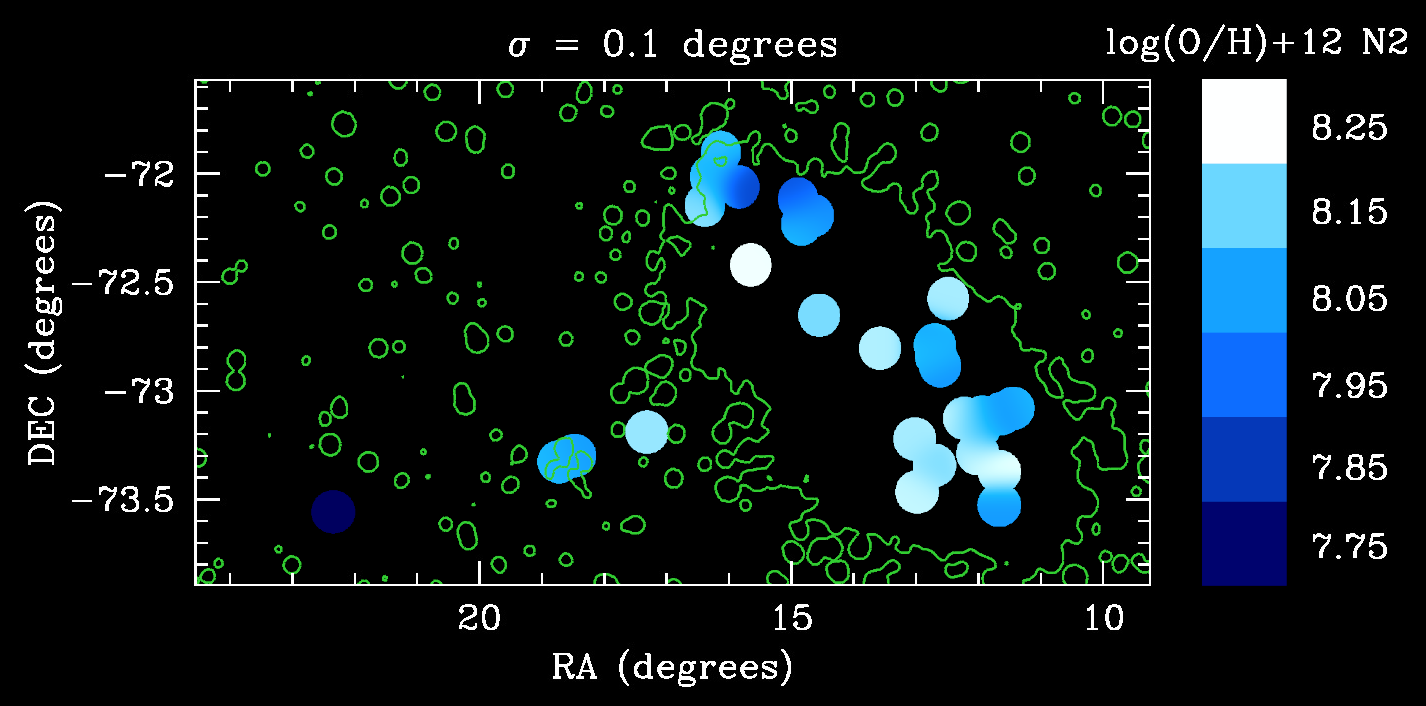}}
  \centerline{\includegraphics[height=7cm]{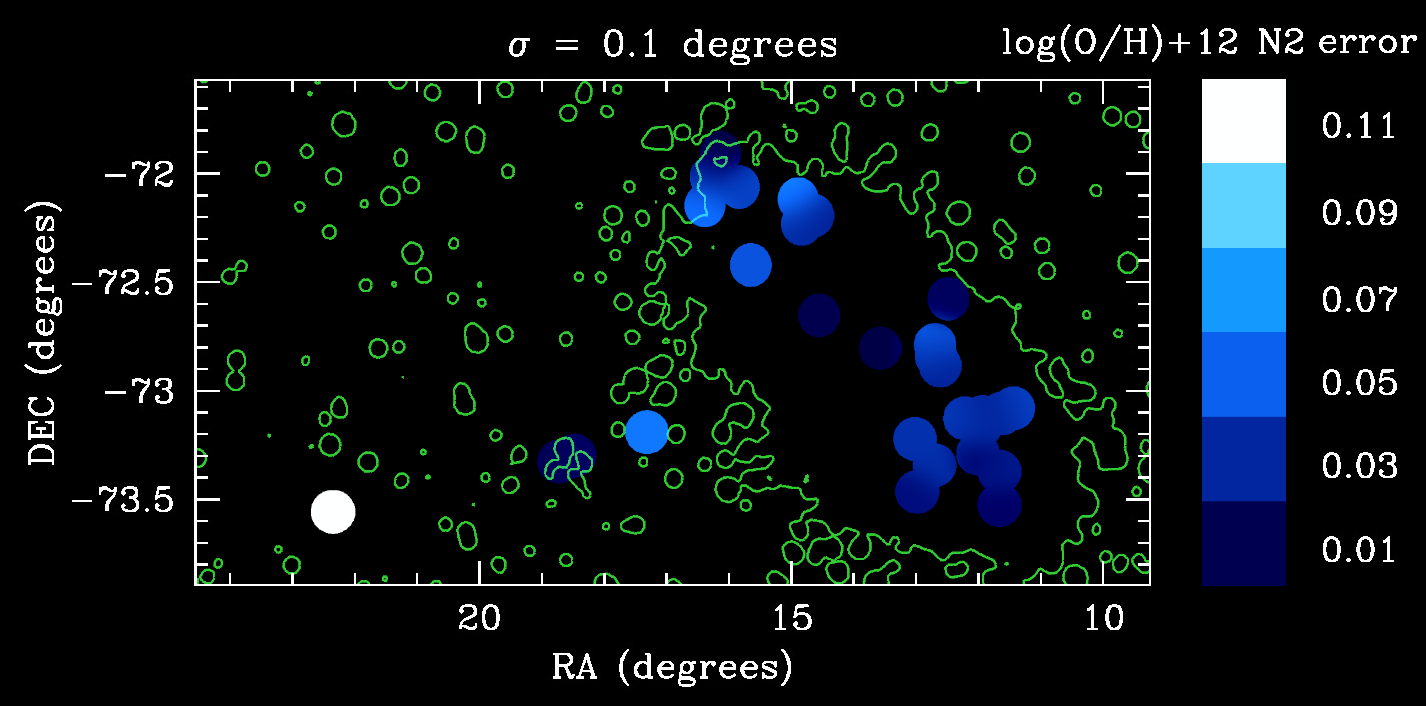}}
  \centerline{\includegraphics[height=7cm]{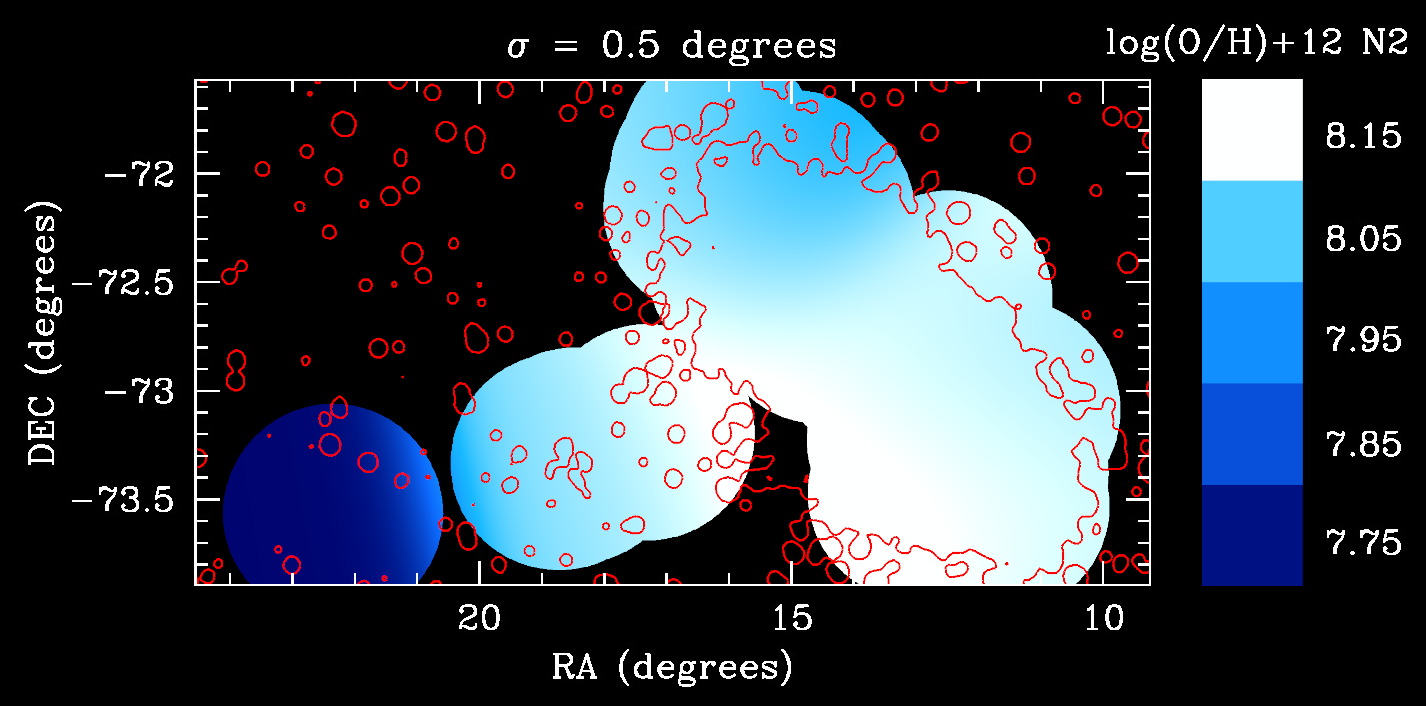}}
  \caption{The SMC \NII\ gas-phase metallicity interpolated with a Gaussian of size 0.1~degrees (top panel) and 0.5~degrees (bottom panel). The middle panel shows the error distribution for the metallicity measurements. In each case the cutoff radius is equal to the Gaussian smoothing size.  The contours are from the SHASSA R band and are used to highlight the location of the SMC main body.}
  \label{fig:metallicity_NII}
\end{figure*}


\begin{table*}
\centering
\caption{The comparison of metallicity values in SMC fields between our work and literature measurements. For our values the error is the random error and does not include the dispersion due to the method used, which can be up to 0.32~dex according to \citet{perezmontero09} or 0.25~dex according to \citet{lopezsanchez12}. The \NII\ were rescaled to O3N2 using the procedure outlined in \citet{kewley08}.}
\label{tab:metallicity_comparison}
\begin{tabular}{ccccccc} 
\hline 
       &        &      &      & Literature  & This Work's & This Work's \\
Source & Object & R.A. & Dec. & log(O/H)+12 & log(O/H)+12 & \NII\        \\
\hline 
\citet{testor01} & N25 & 12.04 & -73.24 & 8.00 $\pm$ 0.05 & 8.250 $\pm$ 0.022 & 8.158 $\pm$ 0.027 \\
\citet{testor01} & N26A-B & 12.03 & -73.25 & 7.98 $\pm$ 0.03 & 8.262 $\pm$ 0.021 & 8.164 $\pm$ 0.026 \\
\citet{reyes15} & N12B & 11.38 & -73.08 & 7.97 $\pm$ 0.03 & 8.045 $\pm$ 0.026 & 8.061 $\pm$ 0.035 \\
\citet{reyes15} & N66 & 14.79 & -72.17 & 7.97 $\pm$ 0.03 & 8.011 $\pm$ 0.024 & 8.003 $\pm$ 0.040 \\
\citet{reyes15} & N81 & 17.32 & -73.20 & 8.09 $\pm$ 0.03 & 8.227 $\pm$ 0.045 & 8.180 $\pm$ 0.058 \\
\citet{reyes15} & N83A & 18.44 & -73.30 & 8.03 $\pm$ 0.03 & 8.096 $\pm$ 0.010 & 8.057 $\pm$ 0.014 \\
\cite{toribiosancipriano17} & N66A & 14.81 & -72.18 & 8.00 $\pm$ 0.02 & 8.021 $\pm$ 0.023 & 8.015 $\pm$ 0.038 \\
\cite{toribiosancipriano17} & N81 & 17.30 & -73.19 & 8.01 $\pm$ 0.02 & 8.227 $\pm$ 0.045 & 8.180 $\pm$ 0.058 \\
\cite{toribiosancipriano17} & NGC~456 & 18.44 & -73.29 & 8.06 $\pm$ 0.05 & 8.096 $\pm$ 0.010 & 8.055 $\pm$ 0.014 \\
\hline 
\end{tabular}
\end{table*}


For our gas-phase metallicity analysis, a smaller sample of 29 fields were considered based on their \Hbeta, \OIIIL, \Halpha\ and \NII$\lambda$6584 measurements. Each line had to have a Gaussian FWHM $> 0.5$\,\AA\ and $< 3$\,\AA, a radial velocity $> 100$\,\kms\ and $< 300$\,\kms, and a signal-to-noise in the line $> 3$ to be considered significant. The gas-phase metallicity was calculated for the SMC fields using the formula from \citet{pettini04},
\begin{equation}
\rm 12 + \log(O/H) = 8.73 - 0.32 \times O3N2 ~,
\end{equation}
where
\begin{equation}
\rm O3N2 = \log \left( \frac{ [OIII]\lambda5007/H\beta } {[NII]\lambda6584/H\alpha} \right) ~.
\end{equation}
This formula for the gas-phase metallicity was used because the ratio of line species both lay in the same observed spectrum, i.e.\ H$\beta$\ and \OIII\ 5007 both lay in the blue spectrum and H$\alpha$\ and \NII\ 6584 both lay in the red spectrum. This meant that any inaccuracies in flux calibration between the blue and red spectrum would not affect the results. Also, because the emission lines in the ratio are fairly close in wavelength, no correction for extinction was necessary. The histogram of the gas-phase metallicities for the SMC fields is shown in Figure~\ref{fig:hist_emission_lines_new_Kewely}. The minimum gas-phase metallicity is 7.87\,dex and the maximum is 8.50\,dex; the mean gas-phase metallicity for the fields is 8.17\,dex (For contrast the mean gas-phase metallicity we find for our LMC observations is 8.33\,dex). Table~\ref{tab:metallicity} in the Appendix lists the flux values for the emission lines used here, along with the positions and gas metallicities for each field.

To display the gas-phase metallicity distribution, a grid in right ascension and declination is constructed containing the SMC. The gas-phase metallicity is calculated for each point in this grid by summing up the metallicity values for each of the fields, using Gaussian weights based on the distance between the field centre and the grid point. The weights, $w$, are thus given by
\begin{equation}
w = \exp(-d^2/(2\sigma^2))
\end{equation}
where $d$ is the distance between the grid point and the field centre and $\sigma$ is the Gaussian kernel size. These weights multiply the metallicities for each field, and the weighted sum is normalised by the sum of the weights to determine the metallicity at the grid point. This yields an interpolated map of the metallicity across the SMC. Points in the grid that are more than one weighting $\sigma$ distance from a pointing are given no value. This prevents extrapolation far from measured fields. 

This interpolated map for the gas-phase metallicity is shown in Figure~\ref{fig:metallicity} with the top panel showing the interpolation using $\sigma = 0.1$~degree and bottom panel using $\sigma = 0.5$~degree. For the $\sigma = 0.1$~degree interpolation, the cutoff means that only a small region around each pointing is shown. Mostly the cutoff is small enough that only the value directly under the interpolation is given weight. However, there are some regions where the fields are close enough together that they affect the values displayed within one cutoff region. The $\sigma = 0.5$~degree interpolation covers regions that do not have \Halpha\ emission. This interpolation matches what one might see for a galaxy at high redshift where individual \HII\ regions are not resolved. It can also be easier to see trends in this smoothed image.

In the $\sigma = 0.1$~degree map, the main body of the SMC shows great variation between neighbouring fields. This variation is within what appears to be a single \HII\ region in the SHASSA image and could be a sign of the disrupted nature of the SMC. This detail is lost in the smoothed $\sigma = 0.5$~degree map, where the larger scale trends show that the central region of the SMC has higher metallicity and is surrounded by lower metallicity wings. The metallicity gradient from the centre to the north of the galaxy is approximately -0.095\,dex/kpc with a shallower metallicity gradient from the centre to the south of the galaxy of approximately -0.013\,dex/kpc. This is consistent with a gas-phase metallicity gradient across the SMC, and corresponds to the stellar metallicity gradient found by many authors \citep{carrera08,dobbie14,piatti15,parisi16,choudhury18,choudhury20,narloch21,parisi22}. The gas-phase metallicity decreases as one moves east along the east-west arm. This suggests that star formation may have started closer to the SMC main body in this arm and then moved outward with time.

The mean oxygen abundance found for the SMC is 8.17\,dex. For comparison, \citet{kurt99} measured values ranging from 7.96 to 8.17\,dex using empirical emission-line diagnostics, \citet{reyes99} measured 7.96\,dex using photoionisation modelling, \citet{garnett99} measured 8.0\,dex from a literature review, \citet{peimbert00} measured 8.15\,dex from collisionally excited lines, \citet{testor01} measured 7.98 to 8.00\,dex, \citet{testor03} measured 7.98\,dex, \citet{penaguerrero12} measured 8.01 to 8.23\,dex from collisionally excited lines, \citet{reyes15} measured 7.96\,dex from comparison to models of the CLOUDY code, and \citet{toribiosancipriano17} measured 7.94 to 8.06\,dex from collisionally excited lines. These measurements often only come from one or two \HII\ regions, but they are remarkably consistent and mostly lower than our value. Some of the difference may be due to the different emission lines used, as often the literature measurements are based on many fainter emission lines. However, \citet{tchernyshyov15} lists a reference photospheric abundance of oxygen metallicity of the SMC of $8.14 \pm 0.08 \pm 0.04$, close to our value. One major factor that may be affecting our results is that the empirical method of \citet{pettini04} that we use has a dispersion of up to 0.32\,dex according to \citet{perezmontero09} or 0.25\,dex according to \citet{lopezsanchez12}. Table~\ref{tab:metallicity_comparison} shows the comparison of individual \HII\ regions where our fields overlap with fields in the literature with specified coordinates.

The gas-phase metallicity we find varies by 0.63\,dex between the highest and lowest measurement. This is quite a large variation compared to other galaxies. Small galaxies are expected to have greater variation, however, as a single star formation event can have a large impact. The amount of variation is not unprecedented, as can be seen in the sample of 49 local field star-forming galaxies from \citet{ho15}.

We also determined the gas-phase metallicity using the \NII\ measurements for the same SMC fields based on the formula from \citet{pettini04},
\begin{equation}
\rm 12 + \log(O/H) = 9.37 + 2.03\,N2 + 1.26\,N2^2 + 0.32\,N2^3 ~,
\end{equation}
where $\rm N2 = log(\NII\lambda 6584/H\alpha)$. A conversion from \citet{kewley08} was used to calibrate these estimates to the above O3N2 estimates and allow direct comparisons:
\begin{equation}
12 + \log(\textrm{O/H}) = -8.0069 + 2.74353 x + -0.093680 x^2
\end{equation}
where $x$ is the original N2 metallicity measurement.

The minimum \NII\ metallicity measurement is 7.73\,dex and the maximum is 8.24\,dex; the mean is 8.10\,dex. Figure~\ref{fig:metallicity_NII} shows smoothed maps created the same way as the maps using the O3N2 calibration. The \NII\ measure of gas phase metallicities preserves the rank-order of the O3N2 gas phase metallicities, but there is a modest systematic bias between the two methods. The O3N2 gas phase metallicity seems to be the better measurement as it does not saturate at high metallicity as the \NII\ seems to do, and it gives smaller formal errors.


\begin{figure}
  \includegraphics[width=\columnwidth]{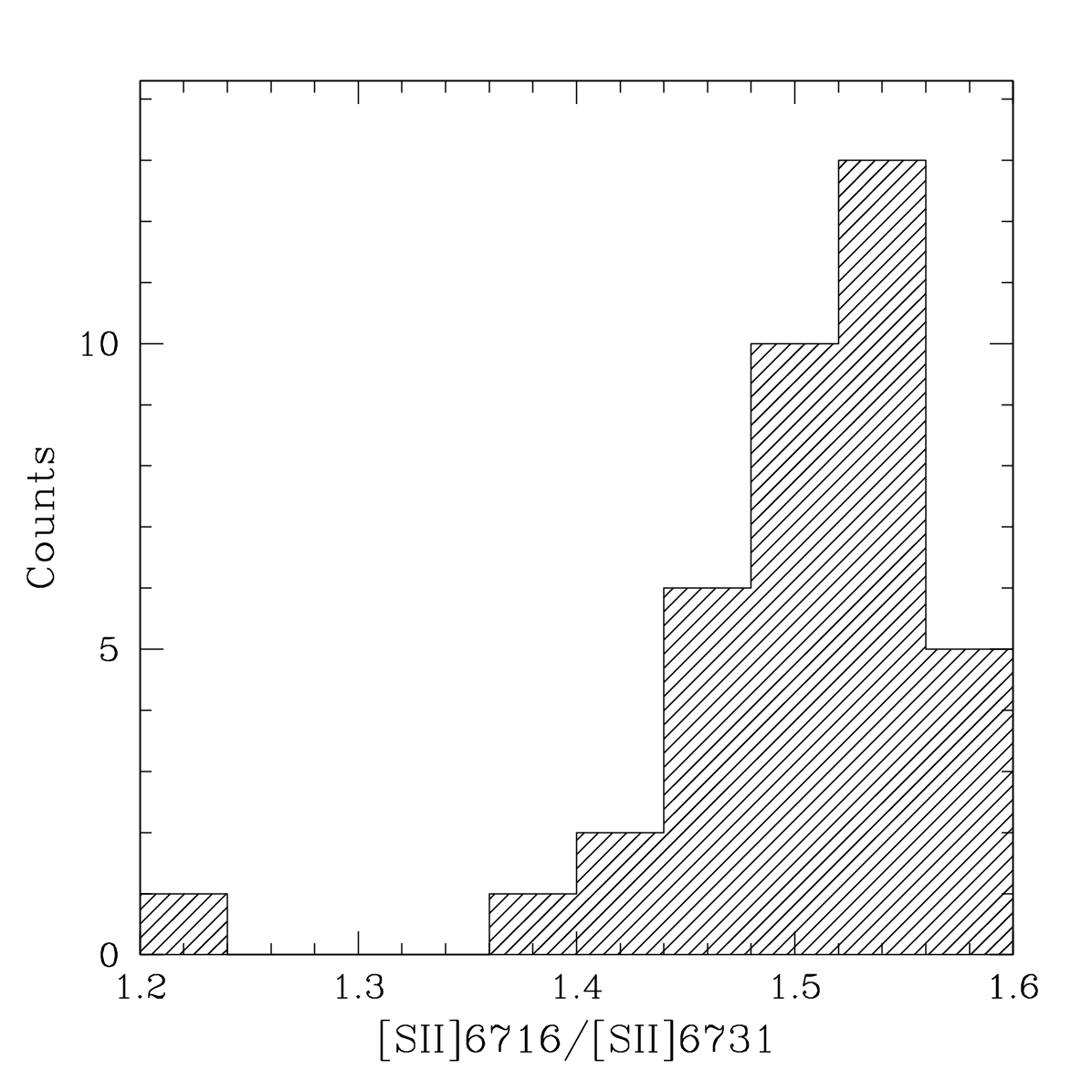}
  \caption{The histogram of the ratio of the emission lines [S{\sc ii}$\lambda$]6716 and [S{\sc ii}]$\lambda$6731 for the fields.}
  \label{fig:hist_emission_lines_electron_density}
\end{figure}


For the fields the [S{\sc ii}] emission lines at 6716~\AA\ and 6731~\AA\ had their fluxes measured to see what could be determined about the electron density from their ratio.  Their ratio is shown in Figure \ref{fig:hist_emission_lines_electron_density}.  The ratio is always above 1.2 (but mostly higher) which lies in the region from electron density 1 to 100 cm$^{-3}$ where the line [S{\sc ii}] ratio to electron density relationship is reasonably flat \citep{sanders16}. Thus all that can be determined is that the electron density for our fields is low. 


\subsection{Extinction}

\label{Extinction}


\begin{figure}
  \includegraphics[width=\columnwidth]{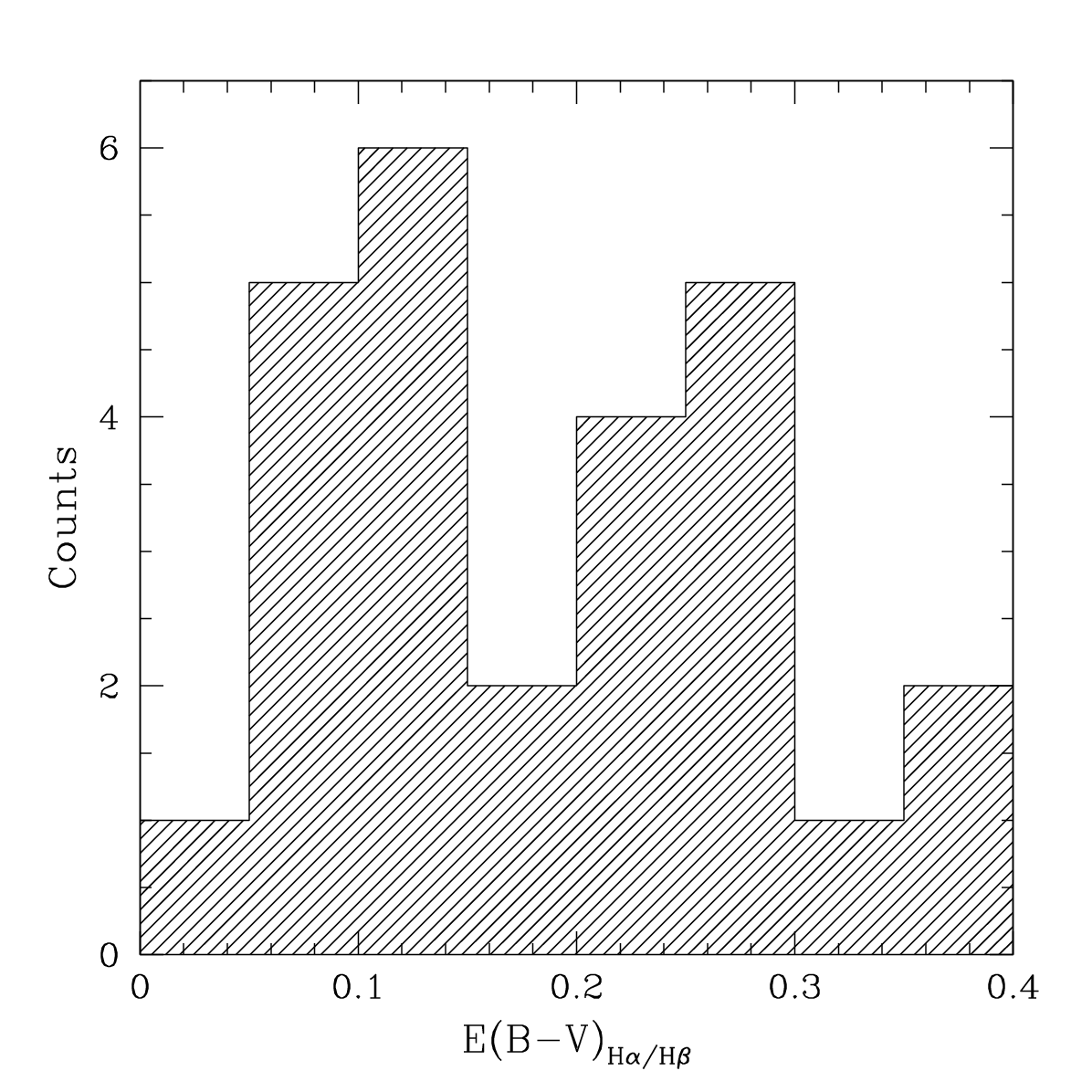}
  \caption{The histogram of $\rm E(B-V)_{H\beta - H\alpha}$ for the 26 significant fields.}
  \label{fig:hist_emission_lines_new_EBV_Halpha_Hbeta}
\end{figure}


\begin{figure*}
  \centerline{\includegraphics[height=7cm]{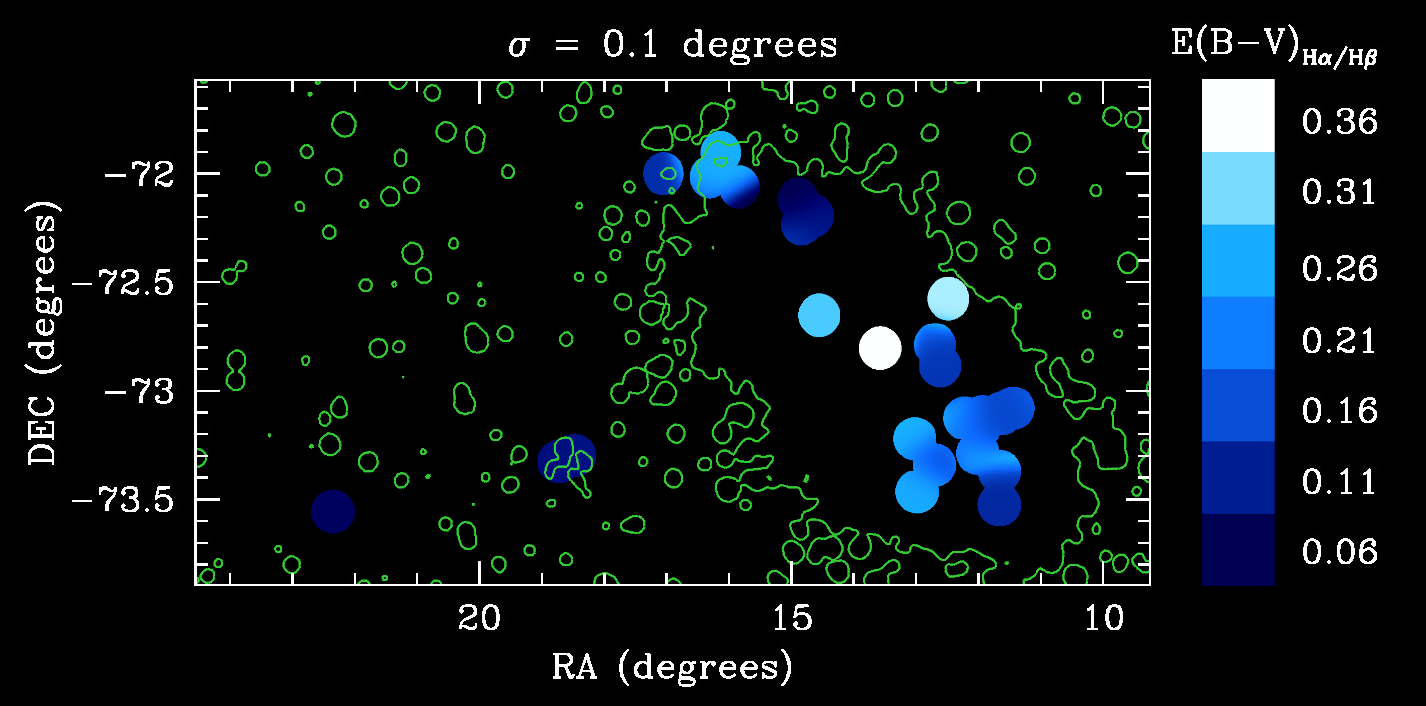}}
  \centerline{\includegraphics[height=7cm]{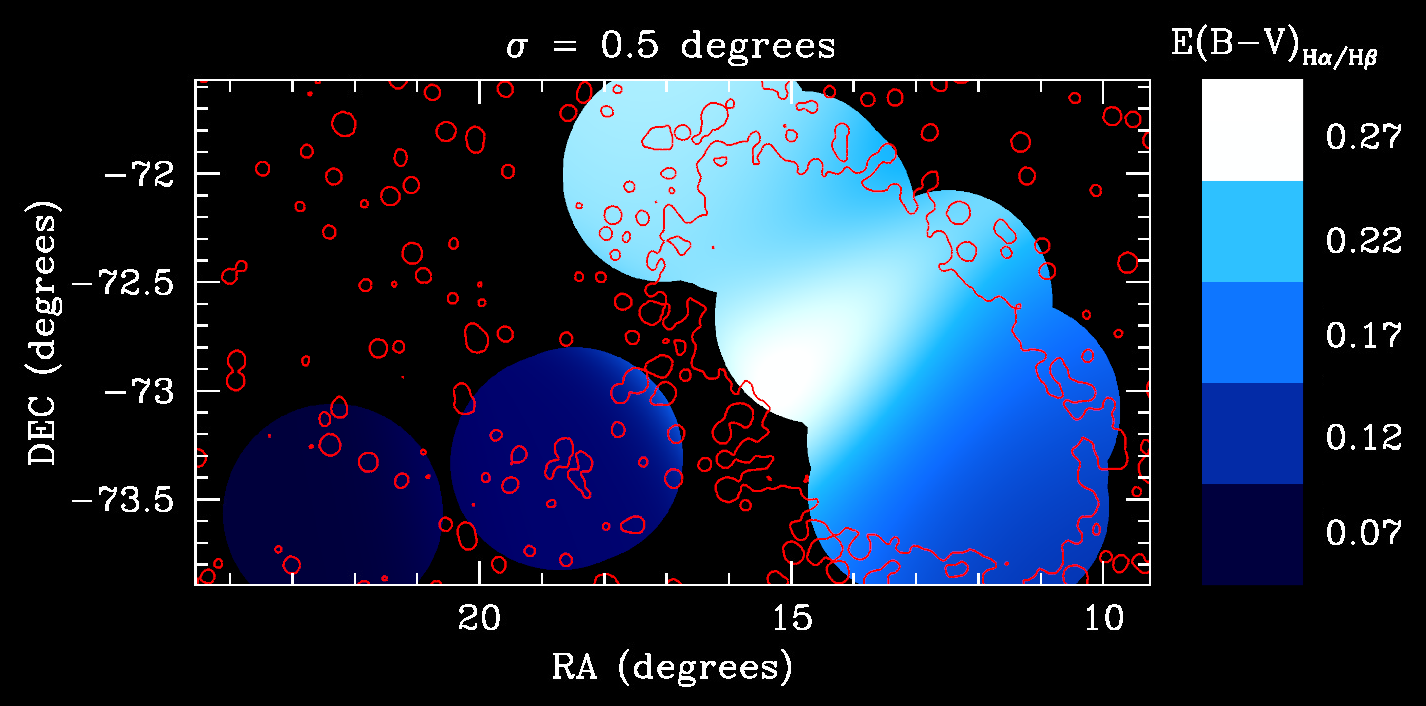}}
  \caption{The interpolated $\rm E(B-V)_{H\beta - H\alpha}$ for the SMC with Gaussian size 0.1~degree (upper panel) and 0.5~degree (lower panel), each with a cutoff radius equal to its Gaussian smoothing size.  The contours are from the SHASSA R band and are used to highlight the location of the SMC main body.}
  \label{fig:extinction}
\end{figure*}


The extinction across the SMC was measured using the ratio of the flux of the \Halpha\ emission line to that of the \Hbeta\ emission line. To calculate the extinction we use
\begin{equation}
\rm E(B-V)_{H\beta - H\alpha} = \frac{-2.5}{k_{H\beta} - k_{H\alpha}} log \frac{(H\alpha/H\beta)_{int}}{(H\alpha/H\beta)_{obs}}
\end{equation}
where $\rm (H\alpha/H\beta)_{obs}$ is the observed flux of ratio of \Halpha\ to \Hbeta\ and $\rm (H\alpha/H\beta)_{int}$ is the intrinsic flux of ratio \Halpha\ to \Hbeta\ (taken to be 2.86 for the case where the B recombination condition has a temperature of T=10,000~K and an electron density of 100~cm$^{-3}$; \citet{osterbrock06}). The values of k$\rm _{H\beta}$ and k$\rm _{H\alpha}$ are taken from SMC bar extinction curve of \citet{gordon03}, using a spline fit to the data giving k$\rm _{H\beta}$ = 3.268 and k$\rm _{H\alpha}$ = 2.171.

To be considered significant for the extinction measurement, the \Halpha\ and \Hbeta\ emission lines of a field had to have a Gaussian FWHM $> 0.5$\AA and $< 3$\AA, a radial velocity $> 100$\,\kms\ and $< 300$\,\kms, and a signal-to-noise in each line flux $> 3$. In addition, the calculation of the extinction had to have error $< 0.1$ in $\rm E(B-V)_{H\beta - H\alpha}$. These conditions left a total of 26 significant fields. The histogram of $\rm E(B-V)_{H\beta - H\alpha}$ for these fields are shown in Figure~\ref{fig:hist_emission_lines_new_EBV_Halpha_Hbeta}. The values for the extinction sample can be found in Table~\ref{tab:EBV} in the Appendix. The average $\rm E(B-V)_{H\beta - H\alpha}$ is 0.185 (in contrast the average extinction for our LMC work was 0.157). The minimum extinction is 0.035 and the maximum is 0.36. Maps, similar to those made for gas-phase metallicity are shown in Figure~\ref{fig:extinction}, the top panel with $\sigma$ = 0.1~degree and the bottom panel with $\sigma$ = 0.5~degree.

In the maps the fields in the eastern arm have low extinction compared to much of the rest of the SMC (starting with a maximum $\rm E(B-V)_{H\beta - H\alpha}$ = 0.91 and then decreasing to 0.68 moving eastward).  The lifetime of a \HII\ region is of order of millions of years, the lifetime of the ionising stars.  According to \citet{jones11} dust lifetimes are of order $10^8$ years from supernova-generated shock waves.  There are very few stars in the east-west arm so one can probably say that the dust has only been produced locally in the star formation region and likely in the recent star formation.  What also hints at the star formation connection is that the metallicity of the gas goes down as one moves eastward along the arm, suggesting only recent star formation in the arm.

The higher extinction values in the centre of the main body of the SMC match the high gas-phase metallicity fields, indicating that these are regions that have seen plenty of star formation, likely over a long period. In the $\sigma$ = 0.5~degree maps there appears to be a gradient from the centre of the main body of the SMC to the ends of the main body similar to the gas-phase metallicity.  This extinction gradient is approximately $-$0.086\,E(B-V)/kpc from the centre going north and shallower, approximately $-$0.0089\,E(B-V)/kpc, going from the centre to the south. 

\citet{gorski20} using red clump stars measured a mean value of the reddening across the SMC of E(B-V)~=~$0.084 \pm 0.013$. Our mean value of 0.185 which is some distance away. However, our E(B-V)$_{H\alpha-H\beta}$ varies from 0.035 to 0.364 across the SMC. In addition, the environment that red clumps stars are found is quite different from the areas of recent star formation that are \HII\ regions. Attempts to match the extinction of individual regions from the literature with our fields were unsuccessful, mainly due to the high uncertainty for individual stars on which the literature measurements were made. 


\subsection{Radial Velocity}

\label{Radial_Velocity}


\begin{figure*}
  \includegraphics[width=\columnwidth]{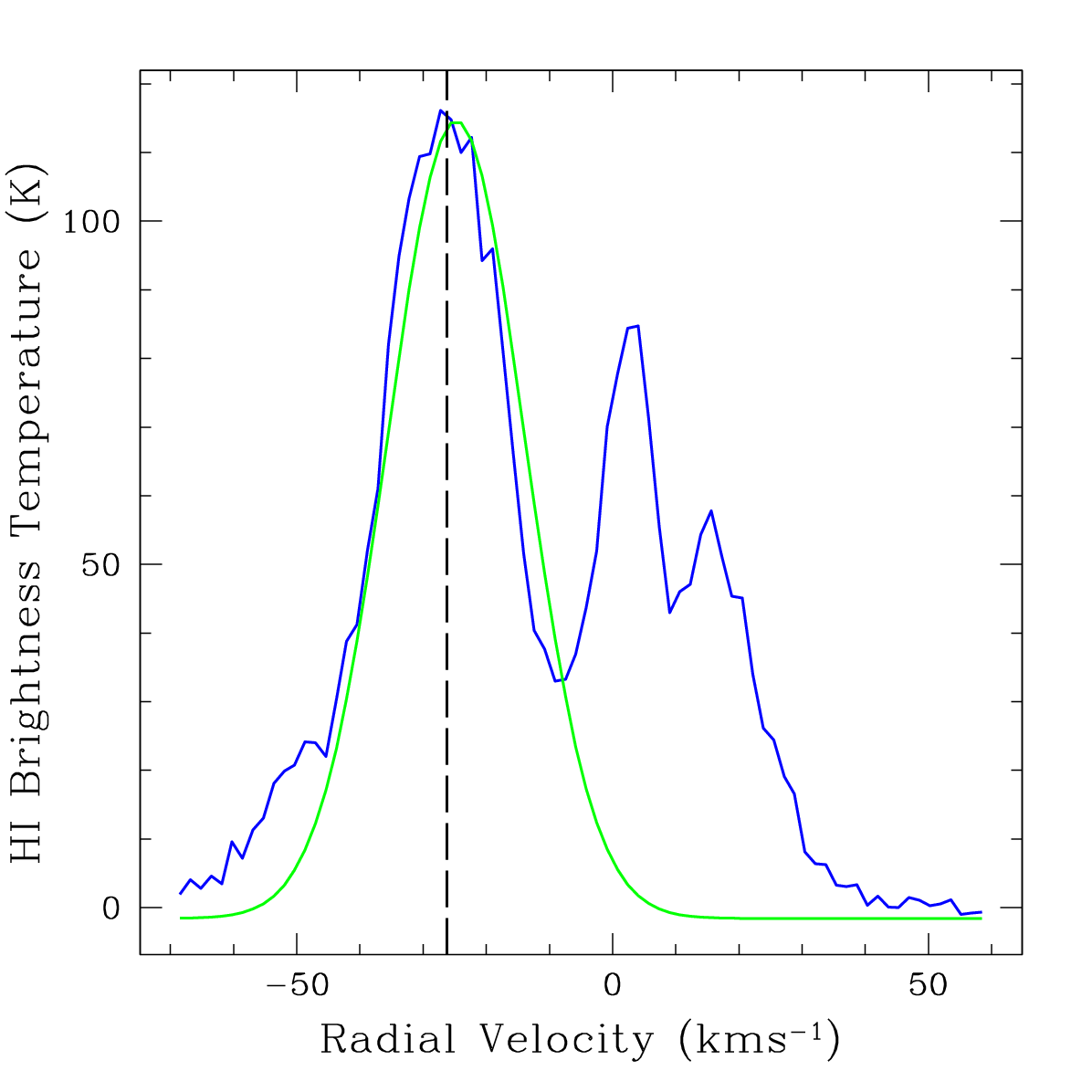}
  \includegraphics[width=\columnwidth]{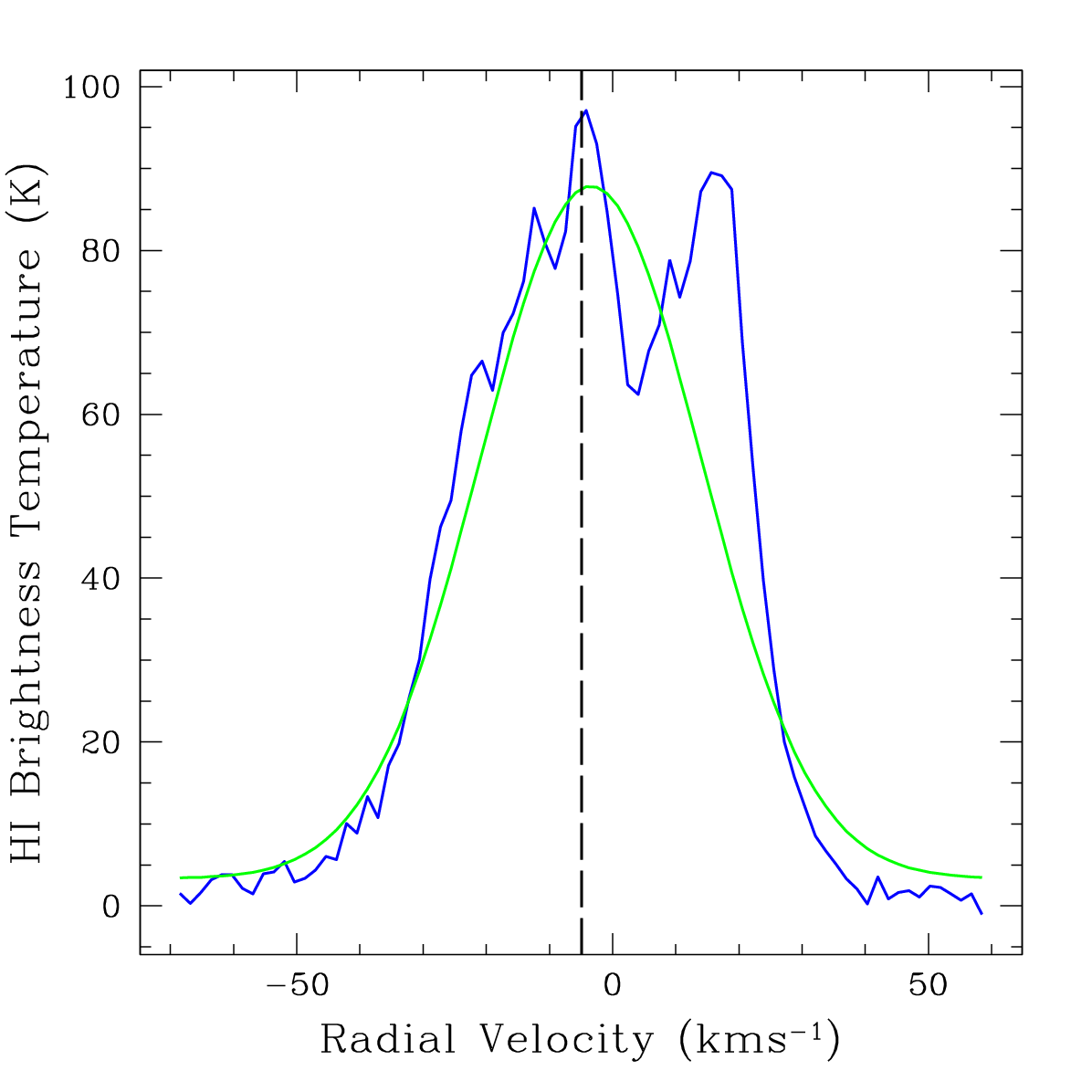}
  \\
  \includegraphics[width=\columnwidth]{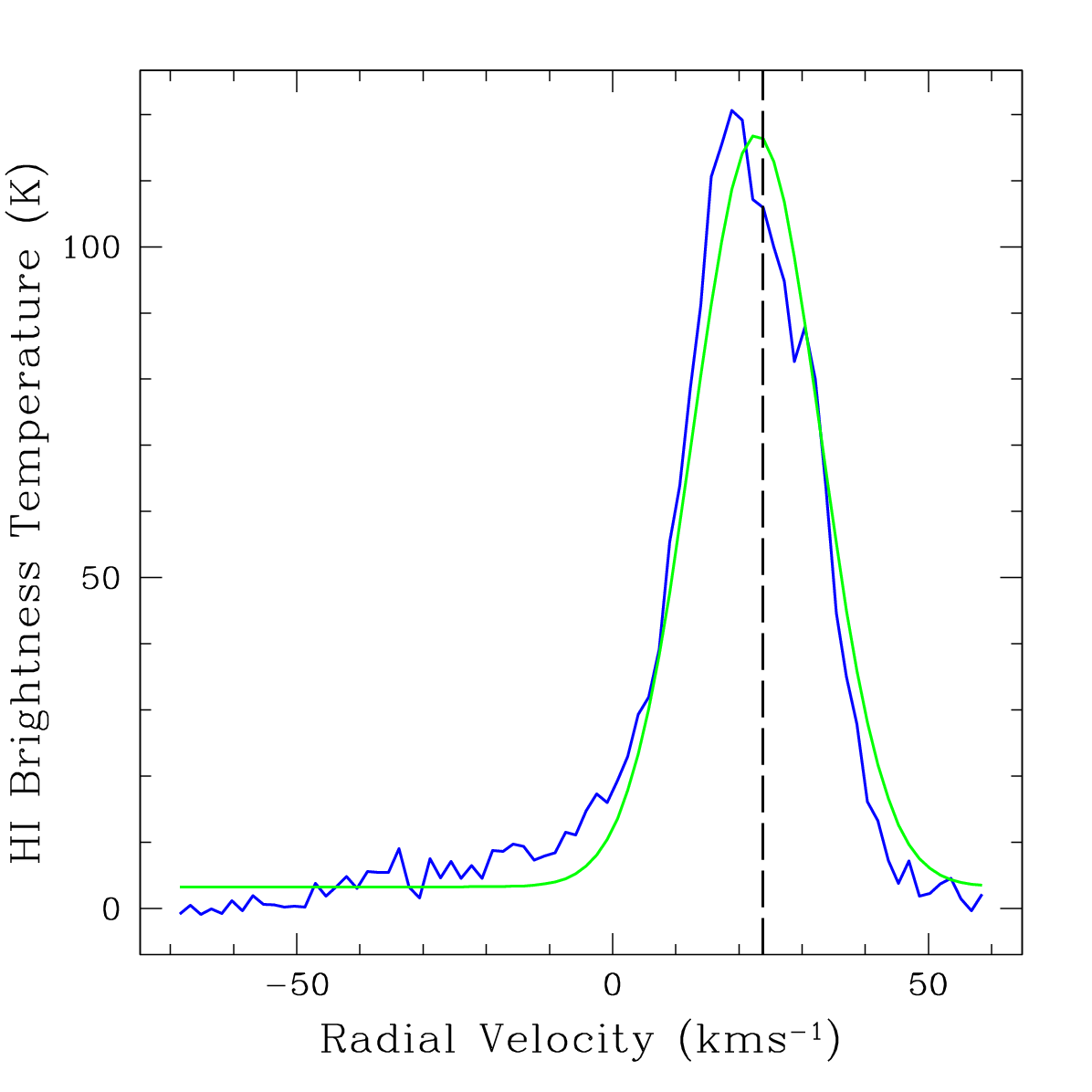}
  \includegraphics[width=\columnwidth]{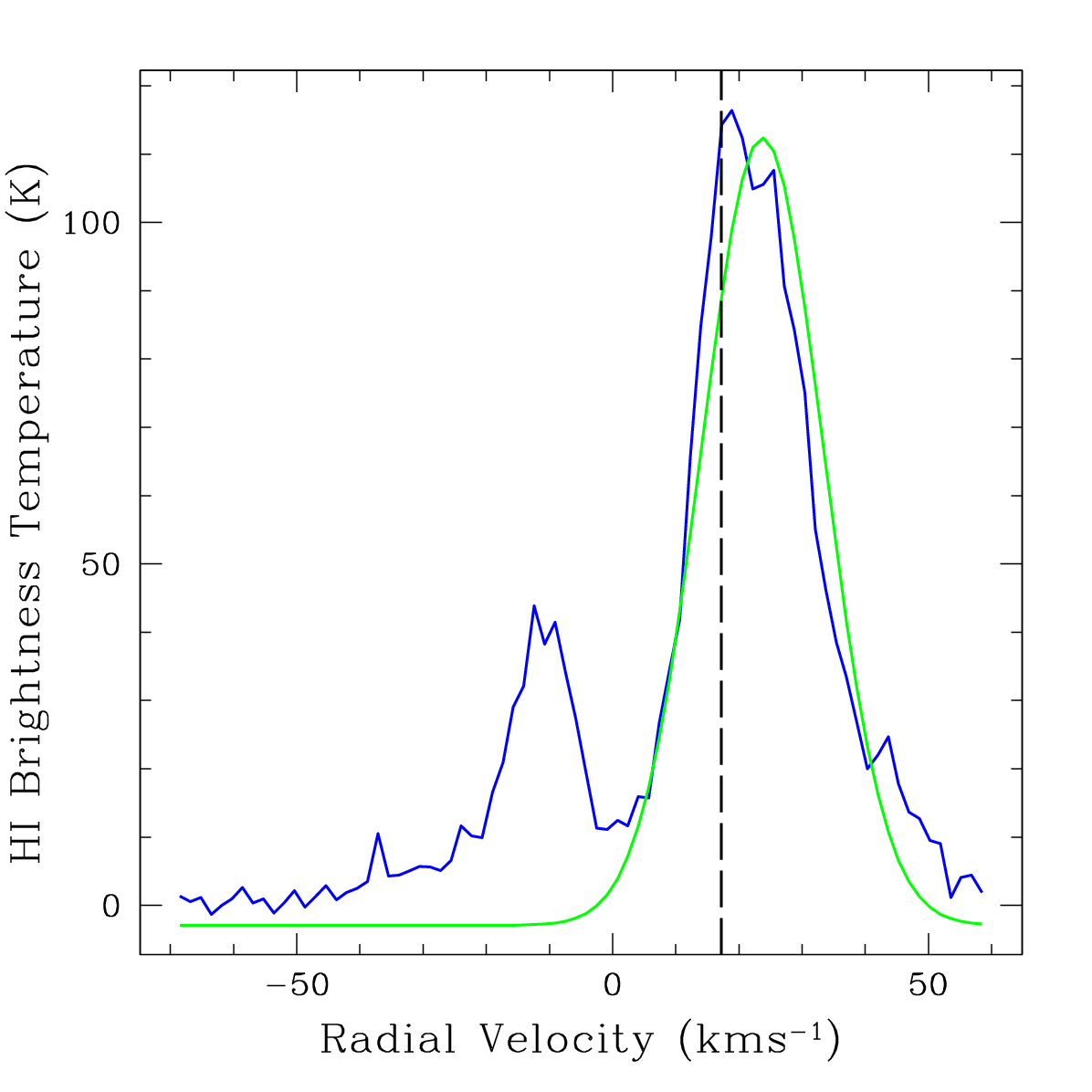}
  \caption{These are four example \HI\ spectra taken at the coordinates of WiFeS fields in the SMC. The \HI\ data are from \citet{staveleysmith95,staveleysmith97,stanimirovic99,stanimirovic04}. The spectra are drawn from a data cube with a 98\,arcsec beam, so around 3 times bigger on a side than the WiFeS field. The zero of radial velocity for the SMC has been set as 157\,\kms\ based on the distribution of our WiFeS fields. The dashed line in each plot is the radial velocity of the \Halpha\ line for the corresponding WiFeS field. The green line is the Gaussian fit to \HI\ emission line linked to the \Halpha\ line. The spectra are organised by increasing Right Ascension. Top left has coordinates R.A.=11\arcdeg658 Dec.=$-$73\arcdeg377, top right R.A.=12\arcdeg702 Dec.=$-$73\arcdeg348, bottom left R.A.=15\arcdeg842 Dec.=$-$72\arcdeg060, and bottom right R.A.=18\arcdeg961 Dec.=$-$73\arcdeg182}
  \label{fig:redshift_HI_raw}

\end{figure*}


\begin{figure}
  \includegraphics[width=\columnwidth]{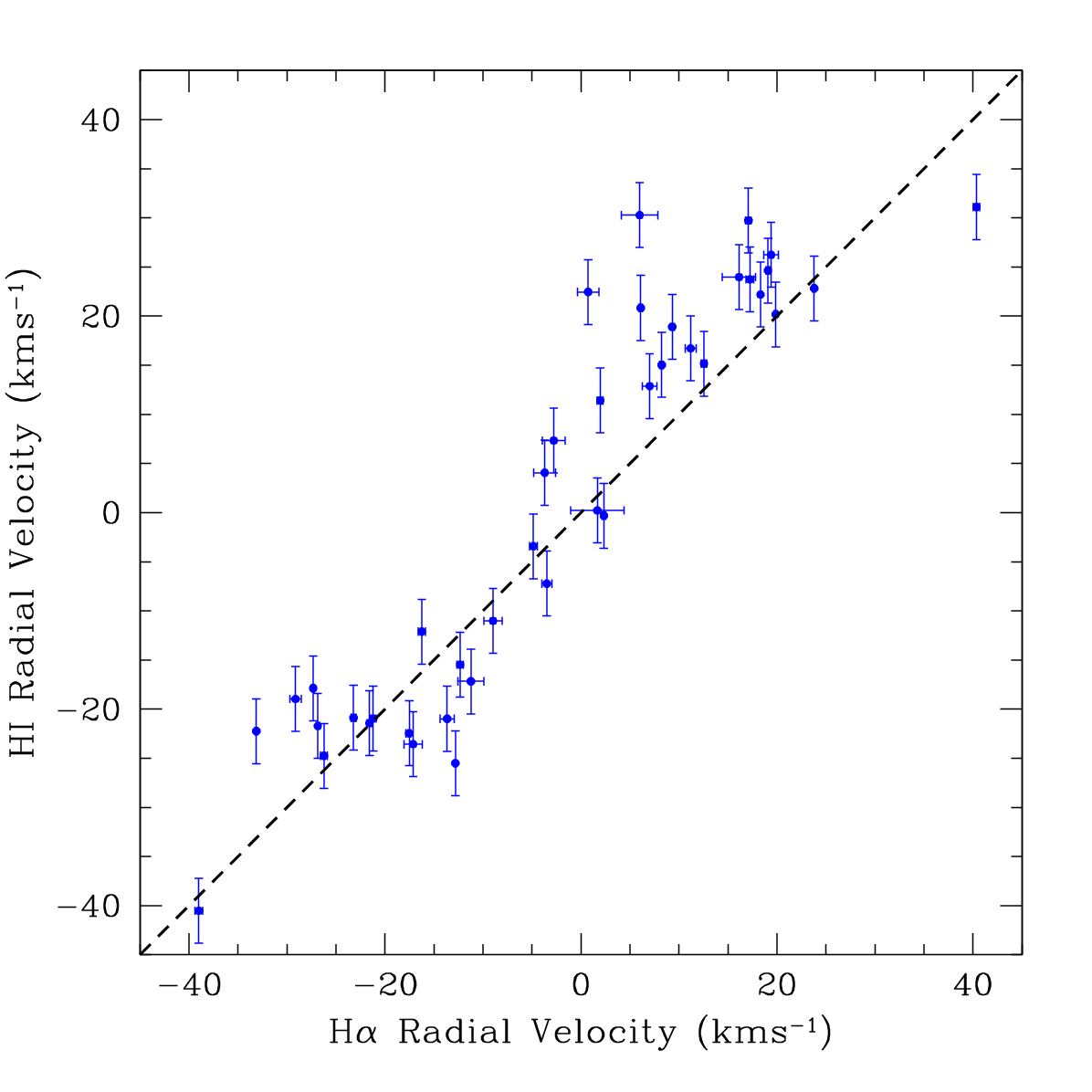}
  \caption{Comparison of the radial velocities of the \Halpha\ emitting gas and the \HI\ gas at the location of the 41 significant fields. The dashed line is 1:1. The zero of radial velocity for the SMC has been set as 157\,\kms\ based on the distribution of our fields.}
  \label{fig:plot_compare_vel_compare}
\end{figure}


\begin{figure*}
  \centerline{\includegraphics[height=7cm]{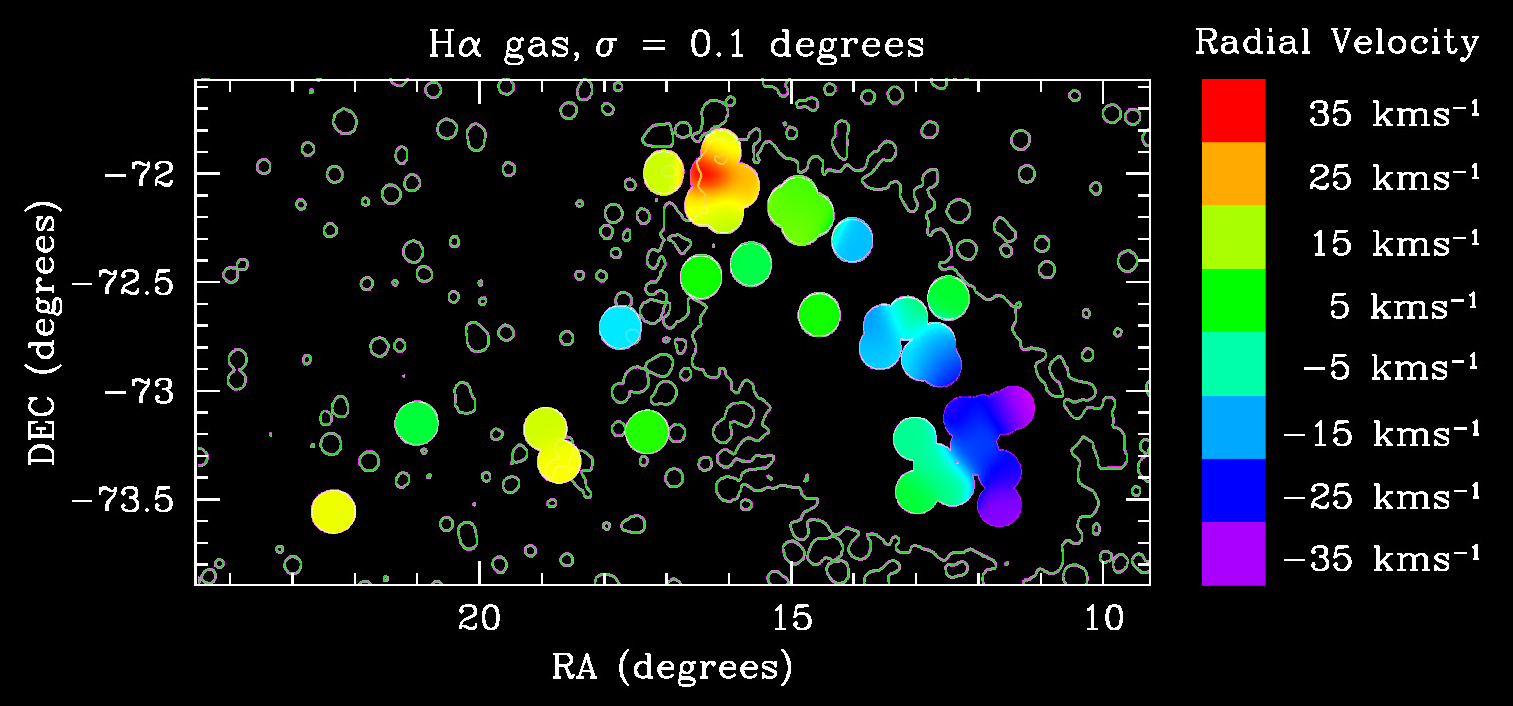}}
  \centerline{\includegraphics[height=7cm]{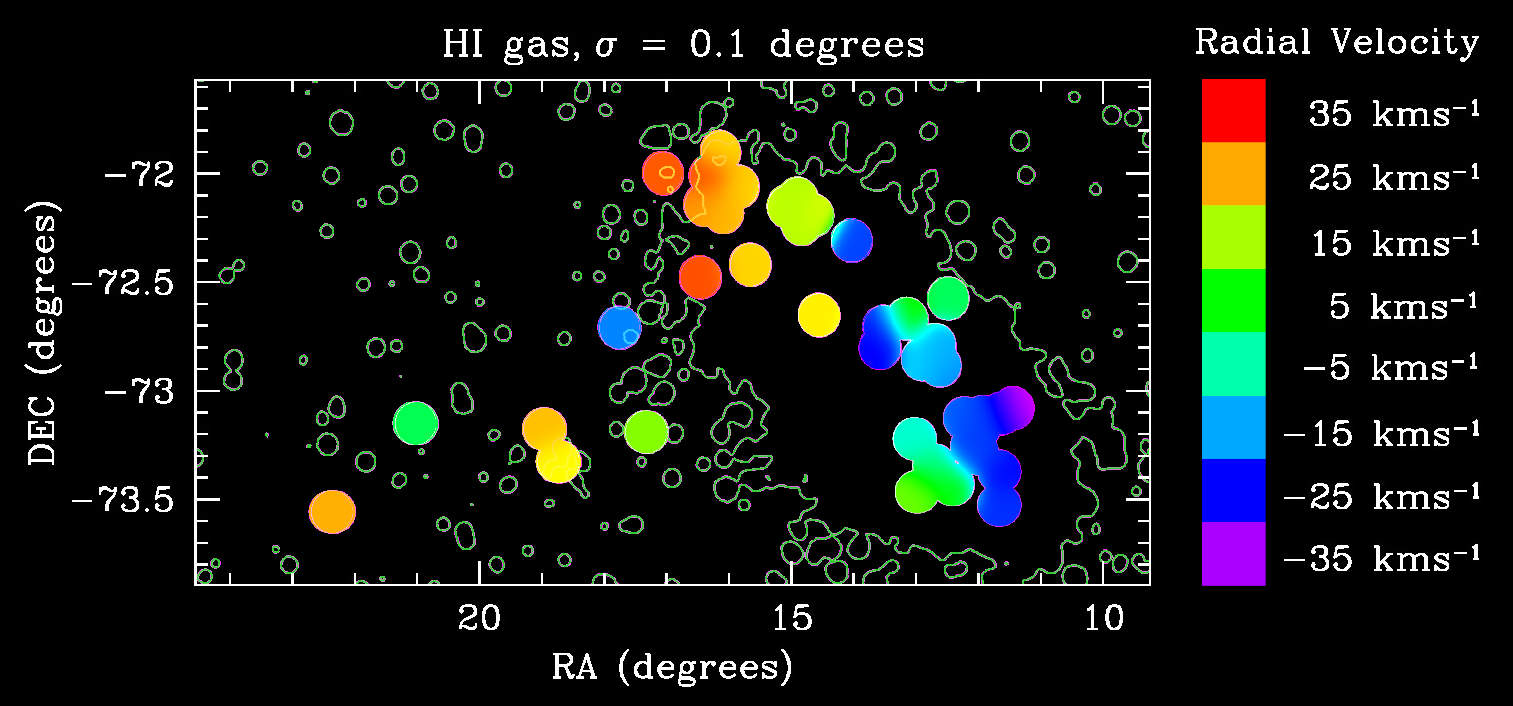}}
  \centerline{\includegraphics[height=7cm]{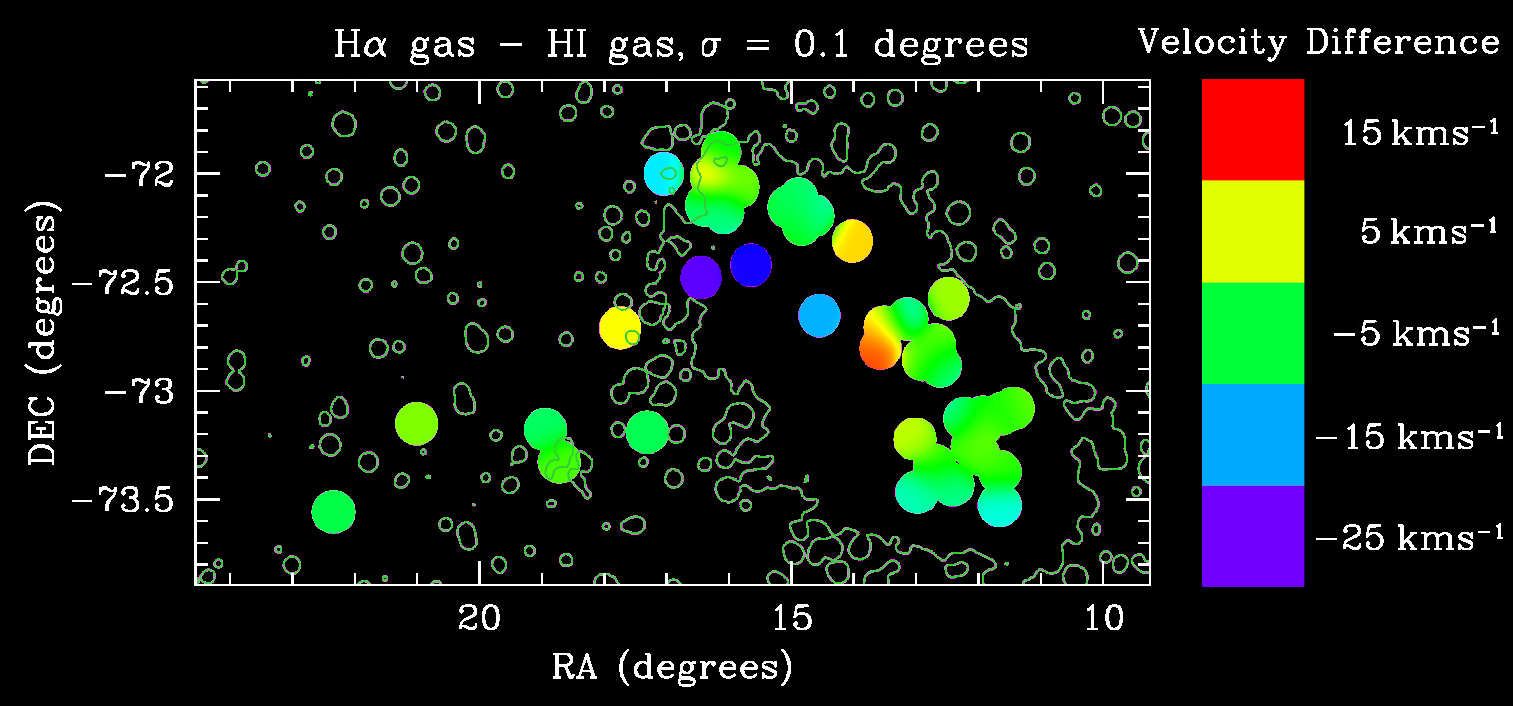}}
  \caption{The radial velocity distribution of the SMC smoothed at 0.1~degree scale. The top panel shows the radial velocities from the \Halpha\ emission interpolated with a Gaussian scale of 0.1~degree and a cutoff size equal to the smoothing size. The middle panel shows the radial velocities from the \HI\ gas interpolated the same way as the \Halpha\ values. The bottom panel shows the difference between the radial velocities of the \Halpha\ emitting gas and the \HI\ gas at this smoothing scale. The contours are from the SHASSA R band and are used to highlight the location of the SMC main body.}
  \label{fig:redshift_0.1}
\end{figure*}


\begin{figure*}
  \centerline{\includegraphics[height=7cm]{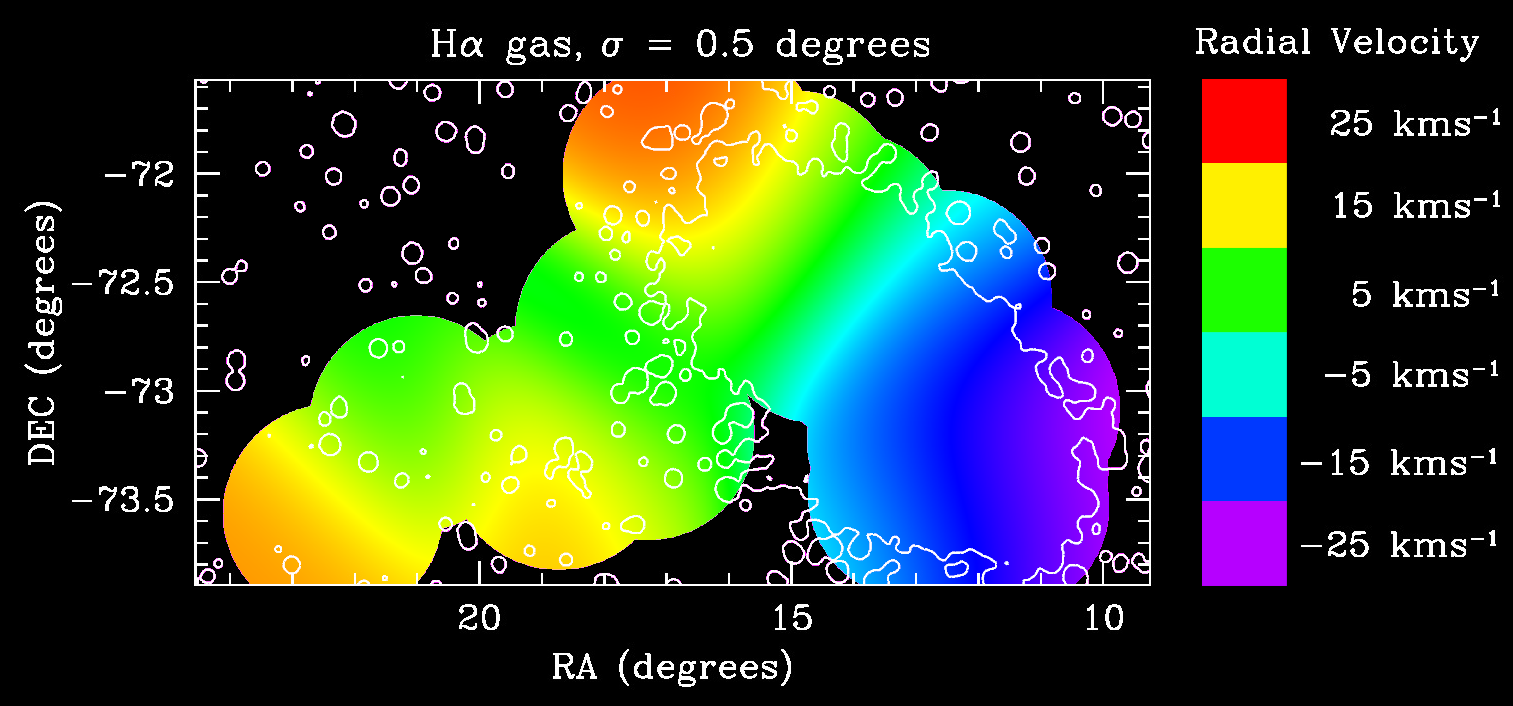}}
  \centerline{\includegraphics[height=7cm]{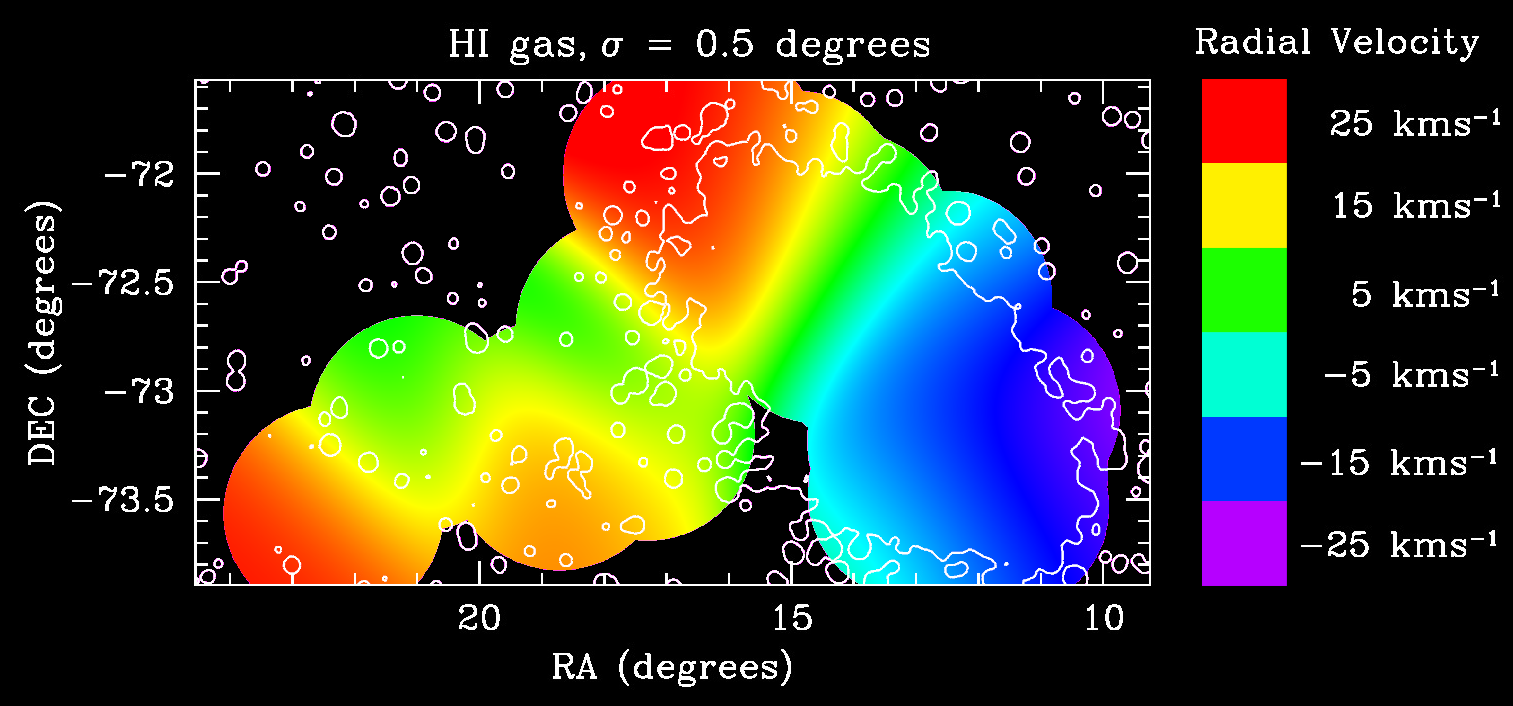}}
  \centerline{\includegraphics[height=7cm]{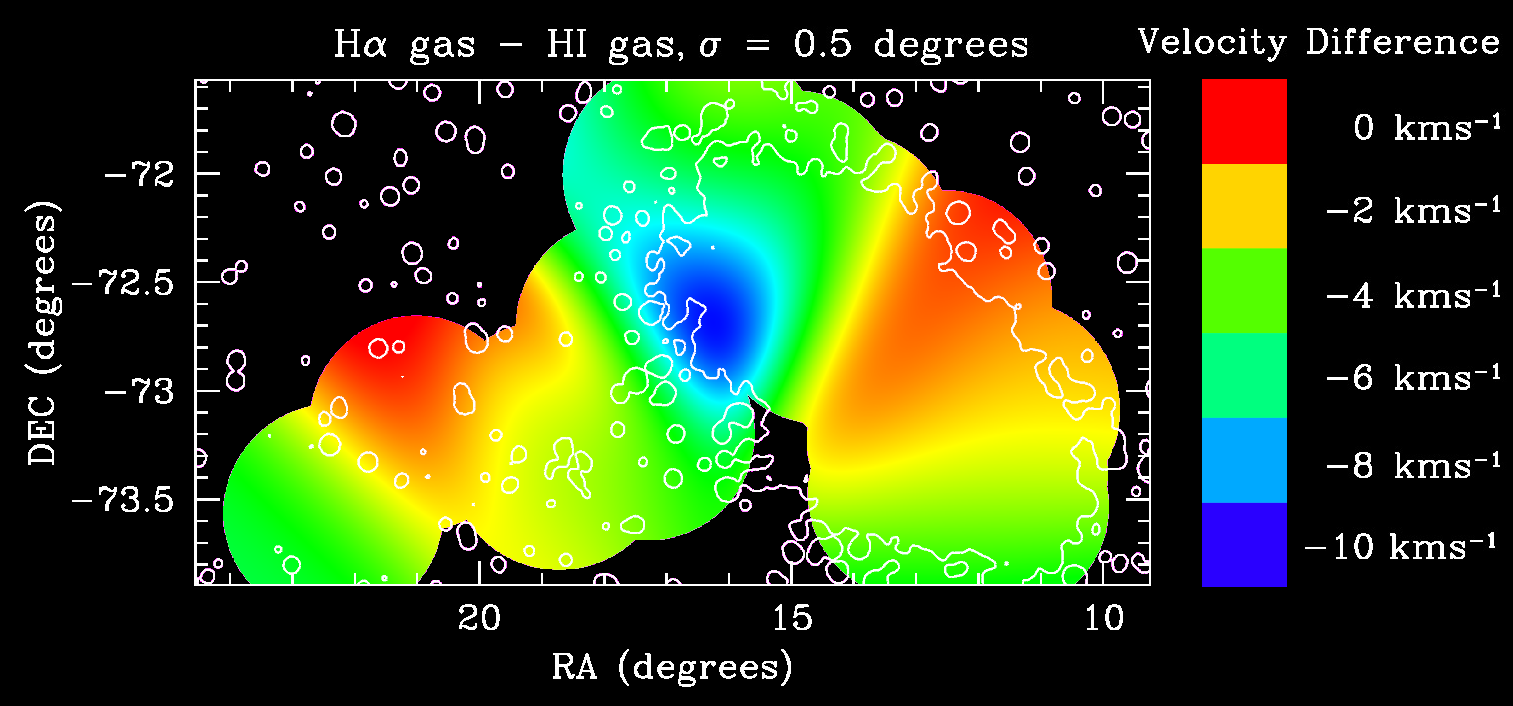}}
  \caption{The radial velocity distribution of the SMC smoothed at 0.5~degree scale.  The top panel shows the interpolated radial velocities from the \Halpha\ emission with a Gaussian scale of 0.5~degree and a cutoff size equal to the smoothing size. The middle panel shows the radial velocities from the \HI\ gas interpolated the same way as the \Halpha\ values. The bottom panel shows the difference between the radial velocities of the \Halpha\ emitting gas and the \HI\ gas at this smoothing scale. The contours are from the SHASSA R band and are used to highlight the location of the SMC main body.}
  \label{fig:redshift_0.5}
\end{figure*}


The radial velocity of each observed field was determined from the \Halpha\ emission line. Measurements of the skylines at 6300.304~\AA\ and 6863.955~\AA\ were used to correct a wavelength offset found in the \Halpha\ line. The skyline at 6300.304~\AA\ gave an offset that varied between 33.5 and 46.5\,\kms\ for different observations; the skyline at 6863.955~\AA\ gave an offset that varied between 30.4 and 41.2\,\kms\ for different observations. A linear interpolation was used to correct for this offset. After this correction there was a systematic error from the wavelength correction of the order of 2\,\kms\ which was deemed negligible given that the spectral pixels corresponded to 20\,\kms\ (the effective velocity resolution is 43\,\kms). Each radial velocity was corrected for the heliocentric velocity at its observed position and date; this varied from $-$7.7 to $-$12.7\,\kms\ for different observations. There are 41 fields with measured \Halpha\ radial velocities.  A systemic velocity of 157\,\kms\ for the SMC was subtracted from the radial velocities derived from the midpoint of our own fields.

We wished to compare the \Halpha\ radial velocities (and later velocity dispersions) from the WiFeS \HII\ regions with the radial velocities (and later velocity dispersions) of the colder \HI\ gas in the same region.  However, there is a problem: the line-of-sight velocity distribution of the \HI\ is often multimodal, with two (sometimes three) velocity peaks in the \HI\ 21~cm emission literature data taken from \citep{staveleysmith95,staveleysmith97,stanimirovic99,stanimirovic04}. Example \HI\ spectra through the \HI\ data cube at the location of WiFeS fields are shown in Figure~\ref{fig:redshift_HI_raw}.  We call each distinct peak a velocity component.  These two components may be the remnants of different unique galaxies that are interacting or one of them may be tidal debris left over from the formation of the SMC \citep{murray24}.  In any case there is only evidence in the \Halpha\ emission line for one velocity component though the spectral resolution is not as fine as the \HI\ spectral resolution.  We therefore may only be seeing the \Halpha\ emission of the component closest to us with the behind component being obscured by stars and dust.  It is possible to determine which \HI\ velocity component is likely to be connected to the \Halpha\ radial velocity system through examination of the \HI\ spectra at each WiFeS location as shown in Figure~\ref{fig:redshift_HI_raw} (the dashed lines in this figure are the \Halpha\ radial velocities).  Because of the multiple velocity component nature of the \HI\ data it was not possible to automate the measurement of the \HI\ radial velocity (and \HI\ velocity dispersion).  What was done instead was use the IRAF task SPLOT to manually fit Gaussians to the \HI\ velocity component linked to the \Halpha\ radial velocity in \HI\ spectra at the location of each WiFeS field.  The green lines in Figure~\ref{fig:redshift_HI_raw} are these Gaussian fits.  From these fits measurements of the \HI\ radial velocity and \HI\ velocity dispersion were made.  The \HI\ radial velocities measured from this method were quite accurate, aligning well with the peaks in the \HI\ spectra. However often the shape of the \HI\ line was not quite Gaussian which meant that the fitted velocity dispersion could only be described as an estimate in these cases.  This method meant that we had measurements at the location of the WiFeS fields but not anywhere else in the SMC.  This becomes relevant later when we compare maps of the \Halpha\ radial velocity (and \Halpha\ velocity dispersion) with that from the \HI\ data. 

The \Halpha\ radial velocity compared to \HI\ 21~cm radial velocity at each field is shown in Figure~\ref{fig:plot_compare_vel_compare} There is generally good agreement at most velocities, although there may be evidence for systematic differences at some velocities that may be structures in the SMC.  To interpret these differences requires examination of the radial velocity maps. For the WiFeS fields, the minimum difference between the \Halpha\ and \HI\ radial velocity is 0.2\,\kms\ and the maximum difference is 24.3\,\kms.

Interpolation maps similar to those for gas-phase metallicity were made of the \Halpha\ radial velocity, with the top panel of Figure~\ref{fig:redshift_0.1} having $\sigma = 0.1$~degree and the top panel of Figure~\ref{fig:redshift_0.5} having $\sigma = 0.5$~degree. In the $\sigma = 0.5$~degree maps, a clear rotation is seen along the main body of the SMC. The eastern arm shows little variation in radial velocity. Looking in detail with the $\sigma = 0.1$~degree maps, the picture is more complex. The southern end of the main body of the SMC has two components, one at \around $-$30\,\kms\ and the other \around 0\,\kms. The dynamics are clearly more complex than a simple rotation along the main body of the SMC and may be signs of the two structures seen in the \HI\ data. 

As we only have \HI\ radial velocities for the WiFeS fields interpolation maps were made the same way as the \Halpha\ radial velocity maps using this data.  The results are shown as the middle panel of Figure~\ref{fig:redshift_0.1} for the $\sigma = 0.1$~degree map and the middle panel of Figure~\ref{fig:redshift_0.5} for the $\sigma = 0.5$~degree map. The difference between the \Halpha\ and \HI\ radial velocity for each field was calculated and the resultant values were then used to make maps.  These can be seen in the bottom panel of Figure~\ref{fig:redshift_0.1} for the $\sigma = 0.1$~degree map and in the bottom panel of Figure~\ref{fig:redshift_0.5} for the $\sigma = 0.5$~degree map.

For many of the regions in common, the \Halpha\ and \HI\ radial velocities in the maps show only small differences. Both \HI\ and \Halpha\ show rotation of the main body of the SMC, as well as a region to the south and east where this rotation does not seem to be along the main body of the SMC.  The most notable difference between the \Halpha\ and \HI\ radial velocity maps is the two fields near the centre of the figure where the \Halpha\ is significantly blue shifted from the \HI.  This shows up in both the 0.1 and 0.5 degree maps though with different velocities due to the smoothing.  The \HII\ regions corresponding to these WiFeS fields may not be connected to the \HI\ gas that lies at the same position.  In the $\sigma = 0.5$~degree map, as expected, much of the detail has been smoothed away.  This includes the two components at the southern end of the main body which have merged into the general rotation of the SMC.

Examination of the stars in the SMC reveals that they lie in a non-rotating spheroid \citep{parisi09,dobbie14b}, in contrast to the \HI\ gas which shows significant rotation (as seen above). \citet{bekki08} explains this difference by suggesting that the SMC had a major merger in the early stages of its formation. This transformed two gas-rich dwarf irregulars into a new dwarf, which consists of a non-rotating spheroidal stellar component and a rotating extended \HI\ disk. The gas ionised by new stars (\Halpha\ emitting gas) generally follows the rotation of the \HI\ gas as expected from the relatively short lifetime of \HII\ regions (\around 10~Myr).  However, we also find important exceptions, highlighting either that these \HII\ regions form in kinematically peculiar regions of the \HI\ disc, or that stellar and supernova feedback can significantly affect the kinematics of the star-forming gas.  At the southern end of the SMC main body, the component to the east that does not share the same radial velocity as the western side may be a structure left over from this merger.  The two components differ by at least 16\,\kms. Indeed, this may be a sign of the two different velocity components seen in the \HI\ data. 


\subsection{Velocity Dispersion}

\label{Velocity_Dispersion}


\begin{figure}
  \includegraphics[width=\columnwidth]{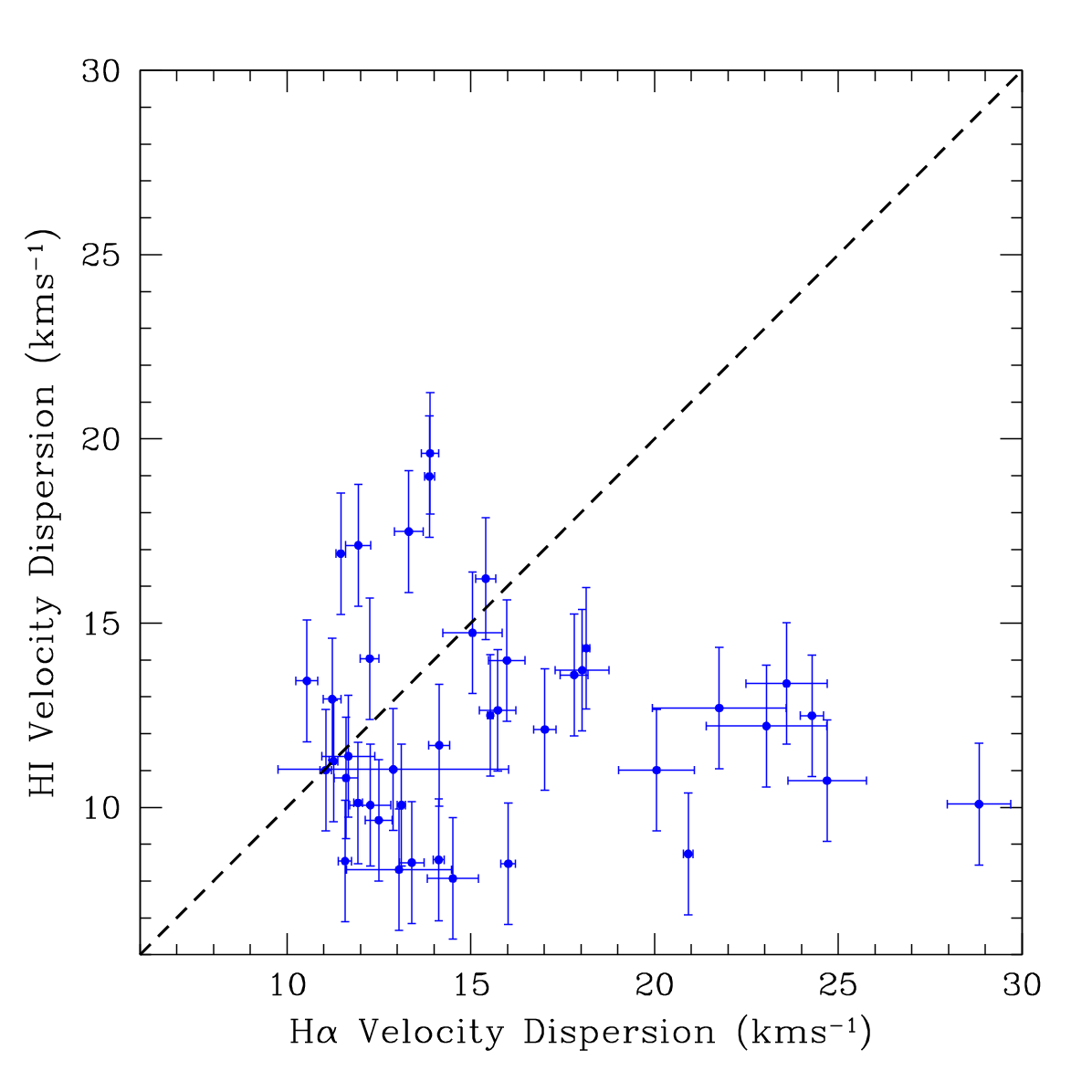}
  \caption{Comparison of the velocity dispersion of the \Halpha\ gas with the \HI\ gas at the location of the 41 significant WiFeS fields. The dashed line is the 1:1 line.}
  \label{fig:plot_compare_vd_compare}
\end{figure}


\begin{figure*}
  \centerline{\includegraphics[height=7cm]{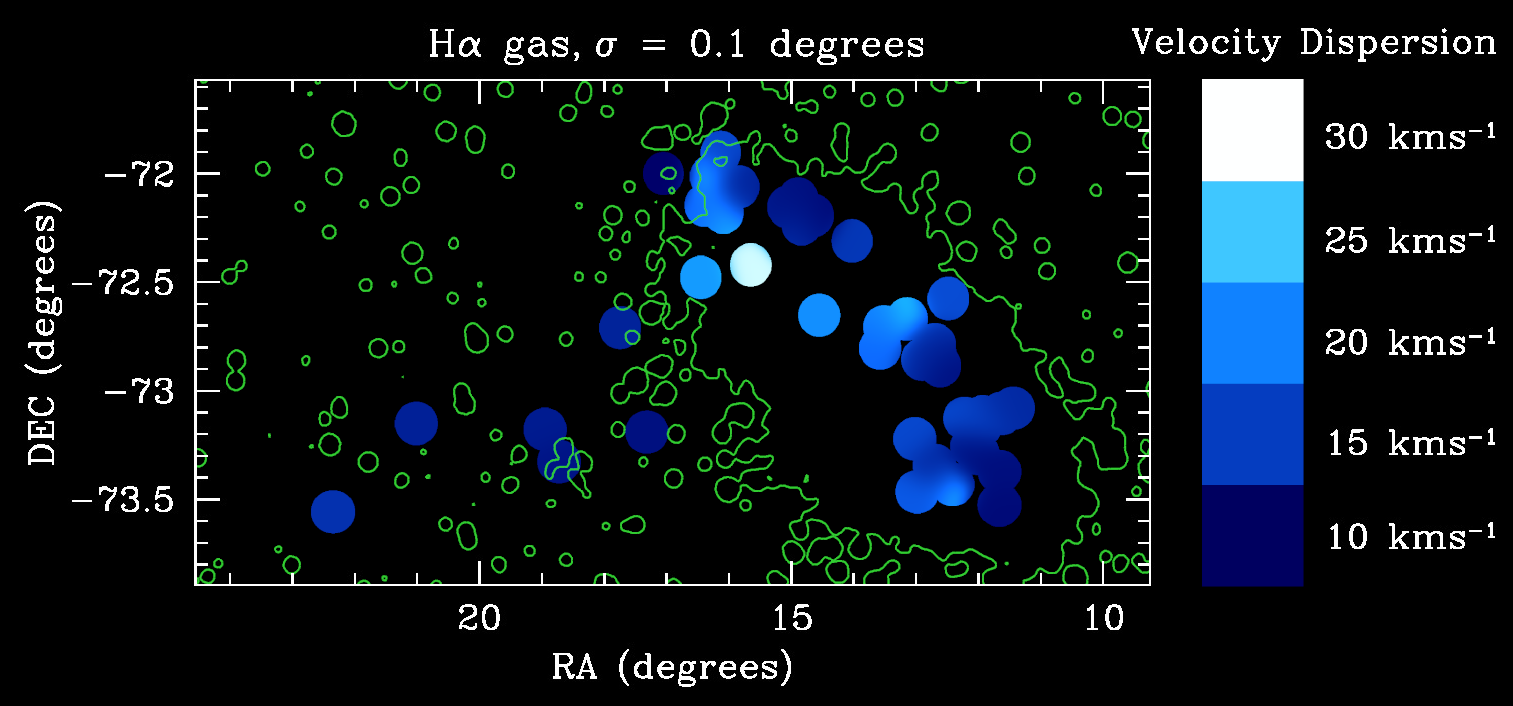}}
  \centerline{\includegraphics[height=7cm]{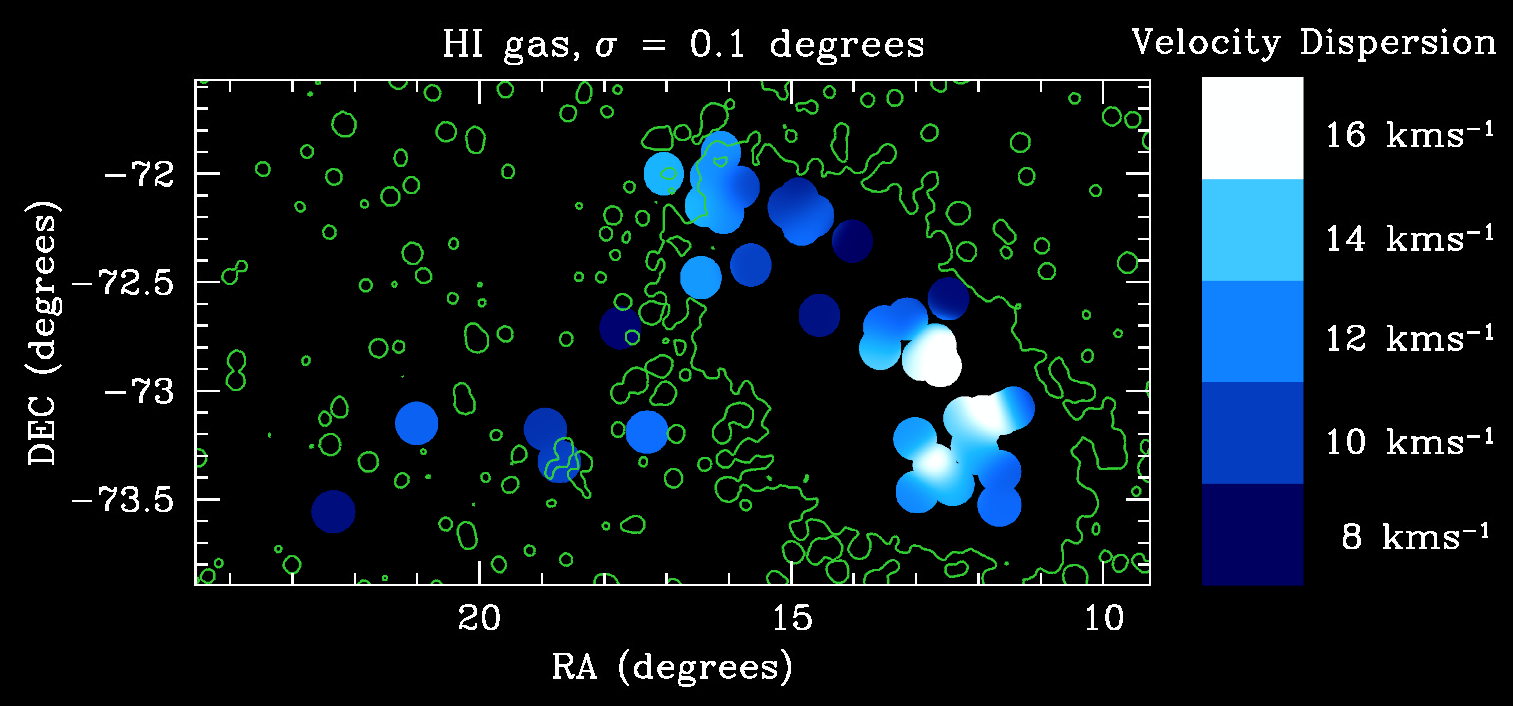}}
  \centerline{\includegraphics[height=7cm]{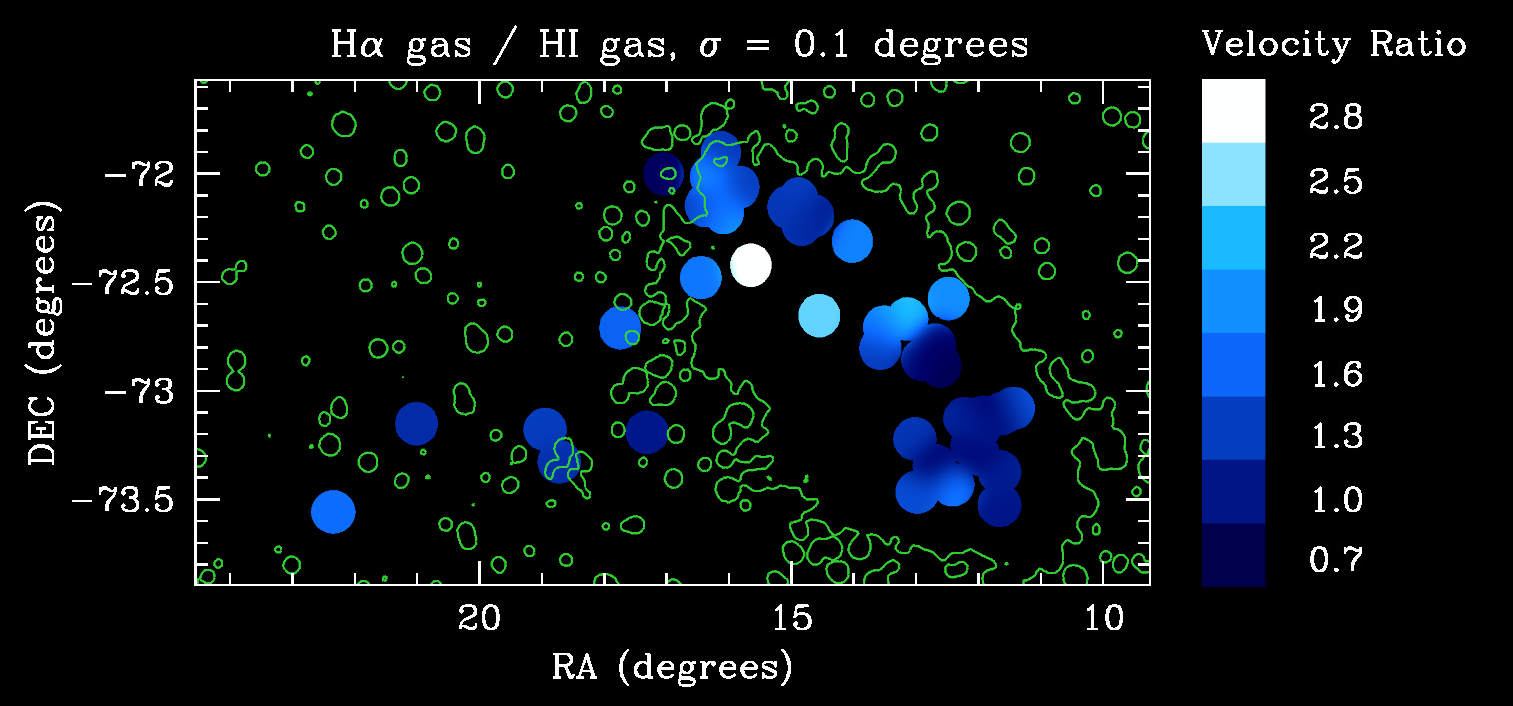}}
  \caption{The velocity dispersion distribution of the SMC smoothed at 0.1~degree scale. The top panel shows the interpolated velocity dispersion from the \Halpha\ emission with a Gaussian scale of 0.1~degree and a cutoff size equal to the smoothing size. The middle panel shows the velocity dispersions from the \HI\ gas interpolated the same way as the \Halpha\ values. The bottom panel shows the ratio of the velocity dispersions for the \Halpha\ emitting gas and the \HI\ gas at this smoothing scale.  The contours are from the SHASSA R band and are used to highlight the location of the SMC main body.}
  \label{fig:velocity_dispersion_0.1}
\end{figure*}


\begin{figure*}
  \centerline{\includegraphics[height=7cm]{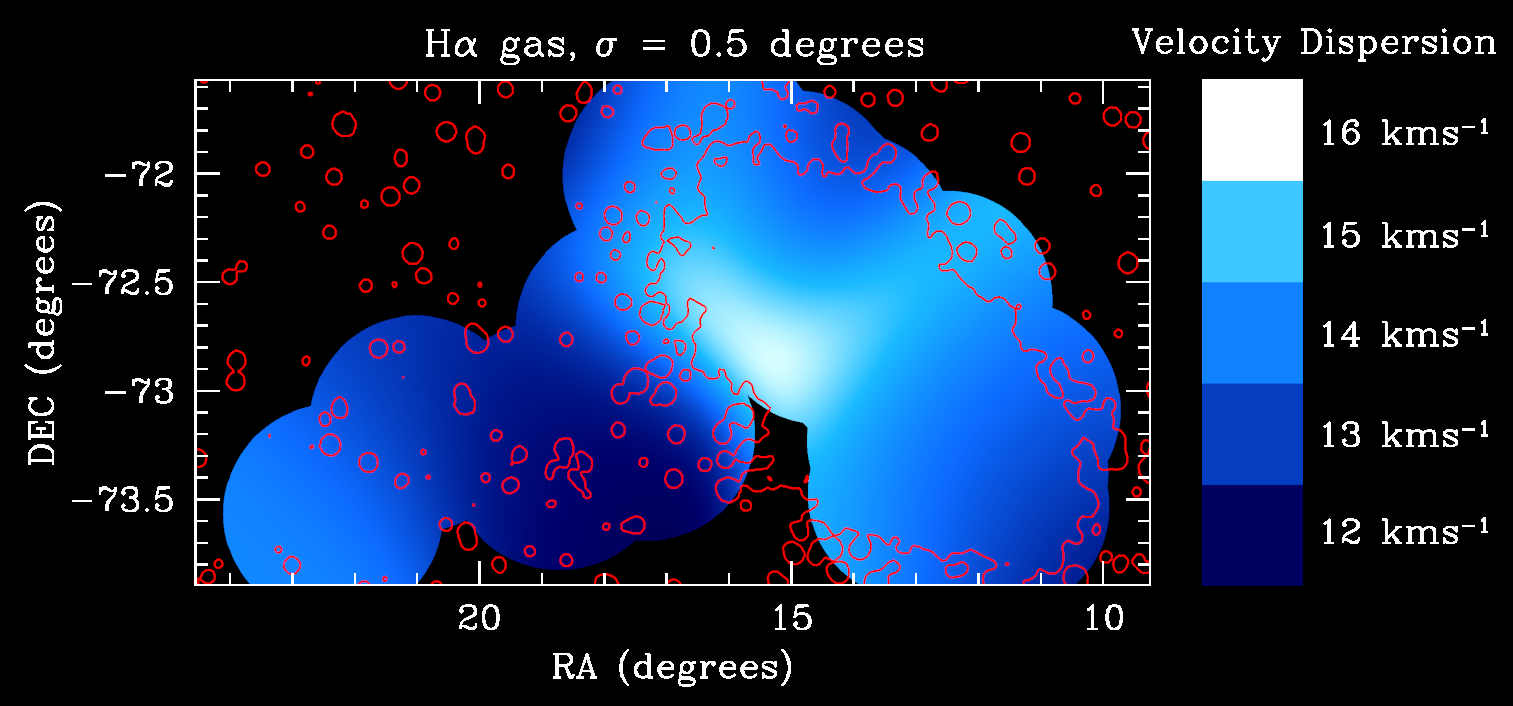}}
  \centerline{\includegraphics[height=7cm]{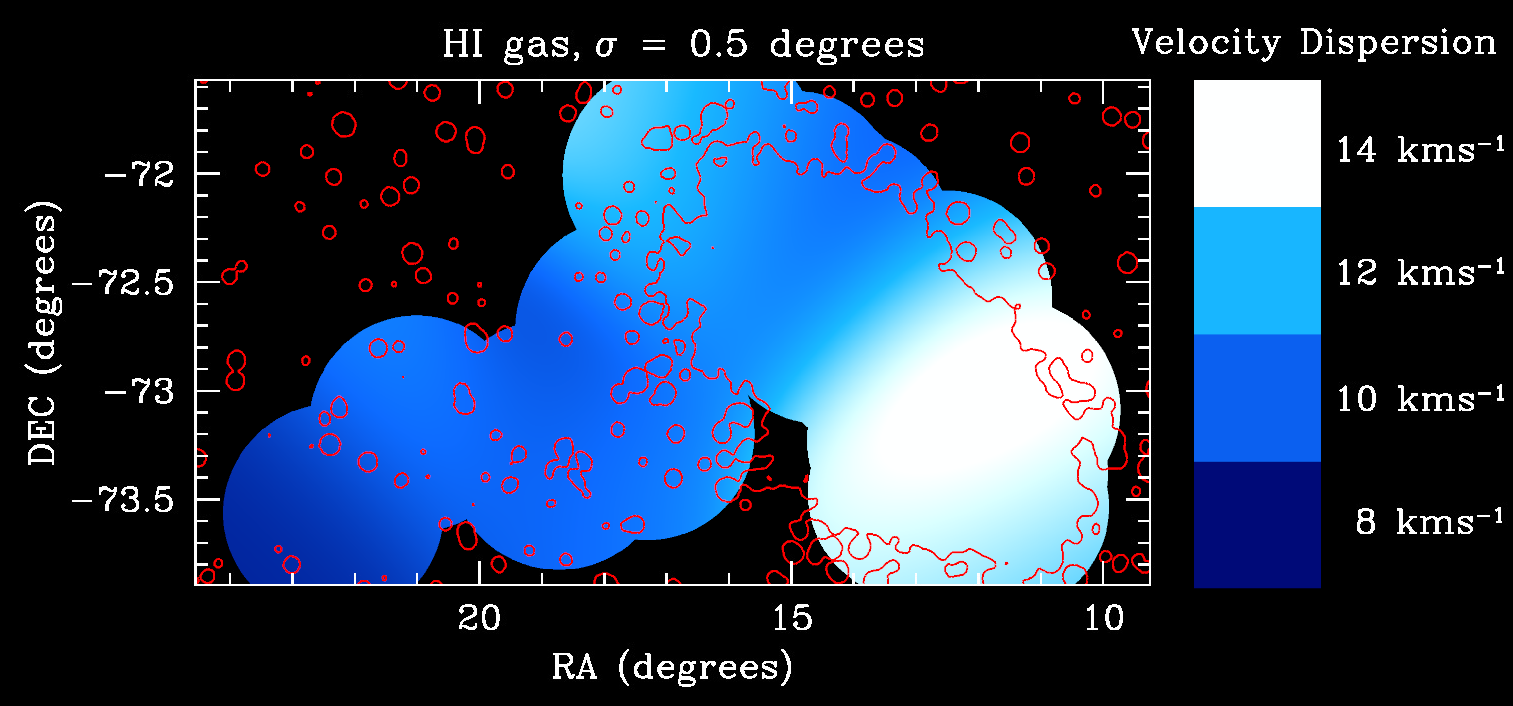}}
  \centerline{\includegraphics[height=7cm]{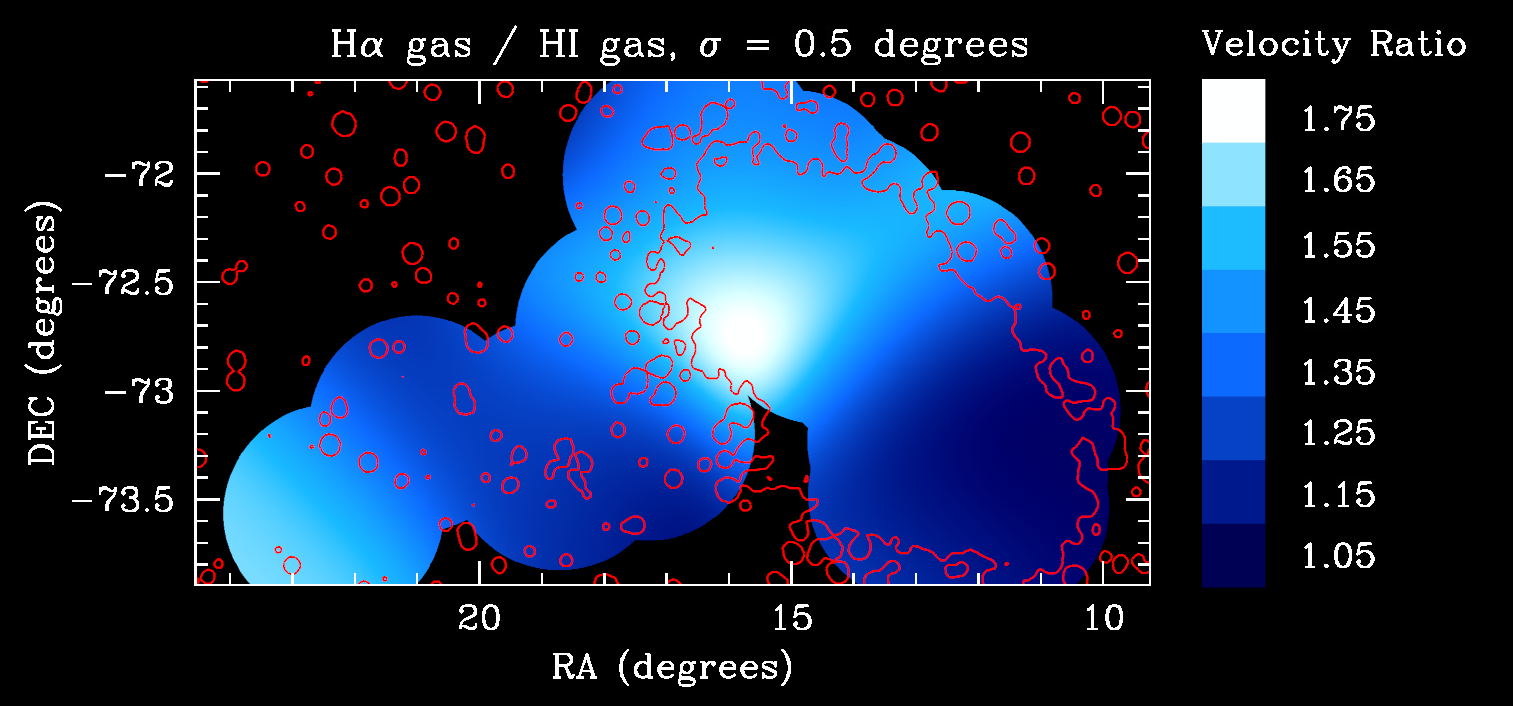}}
  \caption{The velocity dispersion distribution of the SMC smoothed at 0.5~degree scale. The top panel shows the interpolated velocity dispersion from the \Halpha\ emission with a Gaussian scale of 0.5~degree and a cutoff size equal to the smoothing size. The middle panel shows the velocity dispersions from the \HI\ gas interpolated the same way as the \Halpha\ values. The bottom panel shows the ratio of the velocity dispersions for the \Halpha\ emitting gas and the \HI\ gas at this smoothing scale.  The contours are from the SHASSA R band and are used to highlight the location of the SMC main body.}
  \label{fig:velocity_dispersion_0.5}
\end{figure*}


For each of the fields the velocity dispersion has been measured from the width of the \Halpha\ emission line using
\begin{equation}
\rm \sigma_{vd} = \frac{ c }{ 2.355 \lambda (z + 1)} \times \sqrt{\sigma_{H\alpha}^2 - \sigma_{sky}^2}
\end{equation}
with c being the speed of light, $\lambda$ the wavelength of \Halpha\ (6562.8~\AA), $z$ the radial velocity of the emission line, $\rm \sigma_{H\alpha}$ being the Gaussian full width half maximum (FWHM) of the \Halpha\ emission line and $\rm \sigma_{sky}$ being the estimated instrumental resolution of WiFeS at the wavelength of \Halpha\ derived by linear interpolation of the Gaussian FWHM of the strong skylines at 6300.304~\AA\ and 6863.955~\AA. There are 41 fields with measured velocity dispersions. The minimum velocity dispersion is 10.5\,\kms, the maximum 28.8\,\kms, and the mean is 15.6\,\kms.

Figure~\ref{fig:plot_compare_vd_compare} shows, at the position of each pointing, the \Halpha\ line velocity dispersion compared to the \HI\ gas velocity dispersion from \citet{staveleysmith95,staveleysmith97,stanimirovic99,stanimirovic04} measured from the \HI\ spectra as discussed above.  The \Halpha\ velocity dispersion is usually higher than the \HI\ velocity dispersion though not always.  This is similar to what is seen in the LMC \citep{lah24}. The average velocity dispersion for the \HI\ gas is 12.3\,\kms.  The ratio of the \Halpha\ to \HI\ velocity dispersion varies from 0.68 to 2.9.

\Halpha\ velocity dispersion maps have been made similar to those for metallicity. The top panel of Figure~\ref{fig:velocity_dispersion_0.1} shows the \Halpha\ velocity dispersion map with $\sigma = 0.1$~degree and the top panel of Figure~\ref{fig:velocity_dispersion_0.5} shows the \Halpha\ velocity dispersion map with $\sigma = 0.5$~degree. The most noticeable feature is a field with high velocity dispersion in the $\sigma = 0.1$~degree map near the centre of the image. The east arm generally has lower velocity dispersion than the main body of the SMC in the $\sigma = 0.1$~degrees map.  This lower velocity dispersion may be linked to the fact that star formation has only just begun in this region as mentioned above.  The $\sigma = 0.5$~degree map smooths out most of the features, as the velocity range is only 4\,\kms. 

Maps of the velocity dispersion of the \HI\ gas have been created with the same smoothing as the \Halpha\ maps using the \HI\ velocity dispersion at each of the fields. The middle panel of Figure~\ref{fig:velocity_dispersion_0.1} shows the \HI\ velocity dispersion map with $\sigma = 0.1$~degree and the middle panel of Figure~\ref{fig:velocity_dispersion_0.5} shows the \HI\ velocity dispersion map with $\sigma = 0.5$~degree. The ratio between the \Halpha\ velocity dispersion and the \HI\ velocity dispersion at each of the fields was measured and maps were then made.  These can be seen in the bottom panel of Figure~\ref{fig:velocity_dispersion_0.1} for $\sigma = 0.1$~degree and the bottom panel of Figure~\ref{fig:velocity_dispersion_0.5} for $\sigma = 0.5$~degree. 
The \HI\ velocity dispersion is highest at the southern end of the main body of the SMC.  In this region the \Halpha\ and \HI\ velocity dispersions are the closest in value.  

\citet{parisi09} measured a mean velocity dispersion of 23.6\,\kms\ in the 16 SMC star clusters they observed in the SMC. The velocity dispersion of some of the \Halpha\ emitting gas reaches this high (the maximum is 28.8\,\kms), though most is lower; the \HI\ gas velocity dispersion does not reach this high (the maximum is 19.6\,\kms). 


\section{Conclusion}
\label{Conclusion}

Measurements of the gas-phase metallicity across the SMC have been made from multiple fields. The measurements have been interpolated to create a map of the gas-phase metallicity of the SMC. In the $\sigma = 0.1$~degree smoothed map, the main body of the SMC shows great variation in gas-phase metallicity (a variation of 0.63\,dex). In the $\sigma = 0.5$~degree smoothed map, this detail is lost and what appears to be a gradient from the centre of the galaxy can be seen (this varies from $-$0.095\,dex/kpc going northwards to $-$0.013\,dex/kpc going southward). The mean oxygen abundance of 8.17\,dex that we find is generally higher than literature results, except for the value quoted by \citet{tchernyshyov15}. 

Measurements of the extinction across the SMC have been made using the flux ratio of the \Halpha\ emission line and the \Hbeta\ emission line. These have been turned into interpolated extinction maps of the SMC. The eastern arm of the SMC has low extinction (starting with a maximum $\rm E(B-V)_{H\beta - H\alpha}$\,=\,0.91 and then decreasing to 0.68 moving eastward). This suggests that star formation has only begun recently there, as not much dust has been produced. There is high extinction in the centre of the main body of the SMC ($\rm E(B-V)_{H\beta - H\alpha}$\,=\,0.26), which matches the high gas-phase metallicity in that location and indicates that substantial star formation has occurred, likely over a long period. The mean value of the extinction is $\rm E(B-V)_{H\beta - H\alpha}$\,=\,0.185, which is higher than the value found by \citet{gorski20} from red-clump star measurements.

The radial velocity from the \Halpha\ emission line has been measured across the SMC and shows evidence of rotation in the main body of the SMC, with mostly constant radial velocity along the eastern arm. However, there is a region at the southern end of the main body that has a radial velocity different from that expected for the rotation (by at least 16\,\kms), indicating that a more complex velocity structure is present. The \Halpha\ radial velocity has been compared to that from the \HI\ gas. In general there is good agreement between the two emission lines though there are a few notable exceptions. The more complex motion seen at the southern end of the SMC main body may be indications of the two components that make up the \HI\ velocity.

The velocity dispersion across the SMC has been measured from the \Halpha\ emission line.  The eastern arm of the SMC generally has lower velocity dispersion than the SMC's main body (\around 13\,\kms). The \Halpha\ velocity dispersion has been compared to that from the \HI\ emission line and in general the \Halpha\ values are higher (only 9 WiFeS fields out of 41 have higher \HI\ velocity dispersion). The southern end of the main body of the SMC is where the two lines have the closest velocity dispersion.


\section*{Acknowledgements}

We wish to acknowledge the help of Naomi McClure-Griffiths, Lister Staveley-Smith, Mike Bessell, Ian Price, Christopher Lidman, Jon Nielsen and Chris Dowling with this work.

FDE acknowledges support by the Science and Technology Facilities Council (STFC), by the ERC through Advanced Grant 695671 ``QUENCH'', and by the UKRI Frontier Research grant RISEandFALL. 

We make use of data from the Southern H-Alpha Sky Survey Atlas (SHASSA), which is supported by the National Science Foundation. 



\section*{Data Availability}

SHASSA, H$\alpha$\ emission survey data can be found at http://amundsen.swarthmore.edu/SHASSA. The unreduced spectrographic data used for this project is stored in the WiFeS archive. Reduced spectrographic data is available by contacting the authors. \HI\ data came from direct communication with Lister Staveley-Smith.




\bibliographystyle{mnras}
\bibliography{SMC.bib} 


\section*{Appendix}

Table \ref{tab:HI_regions} contains various measured values for significant \HII\ regions for ease of comparison.  Tables~\ref{tab:Halpha}, \ref{tab:metallicity}, and \ref{tab:EBV} contain the data for each field used in this work.


\begin{table*}
\centering
\caption{Various measured values for significant \HII\ regions within the SMC.}
\label{tab:HI_regions}
\begin{tabular}{rllll} 

\hline
\HII\ region                        & NGC 346           & NGC 395           & HN 30 \\
\hline

R.A.                                & 14.6446           & 16.3108           & 11.6188 \\
Dec.                                & $-$72.1928        & $-$72.0133        & $-$73.1075 \\
Gas phase metallicity               & $8.030 \pm 0.018$ & $8.159 \pm 0.023$ & $7.993 \pm 0.015$ \\
$\rm E(B-V)_{H\beta - H\alpha}$     & $0.091 \pm 0.041$ & $0.143 \pm 0.054$ & $0.139 \pm 0.027$ \\
\Halpha\ radial velocity (\kms)     & $17.9 \pm 0.2$    & $48.7 \pm 0.3$    & $-17.5 \pm 0.2$ \\
\HI\ radial velocity  (\kms)        & $18.9 \pm 3.3$    & $31.1 \pm 3.3$    & $-21.7 \pm 3.3$ \\
velocity difference (\kms)          & $1.0 \pm 3.3$     & $17.6 \pm 3.3$    & $4.2 \pm 3.3$ \\
\Halpha\ velocity dispersion (\kms) & $11.1 \pm 0.2$    & $24.3 \pm 0.3$    & $13.9 \pm 0.1$ \\
\HI\ velocity dispersion  (\kms)    & $11.0 \pm 1.65$   & $12.5 \pm 1.65$   & $19.0 \pm 1.65$ \\
velocity dispersion ratio           & $1.01 \pm 0.15$   & $1.94 \pm 0.26$   & $0.732 \pm 0.062$ \\

\end{tabular}
\end{table*}



\begin{table*}
\centering
\caption{The \Halpha\ fluxes, radial velocities and velocity dispersions of the 41 significant WiFeS fields. R.A.\ and Dec.\ are in degrees, fluxes in erg\,s$^{-1}$\,cm$^{-2}$\,\AA\!\!$^{-1}$, and radial velocities and velocity dispersions in \kms. The error in the radial velocity is the random error; there is a systematic error from the wavelength correction that can be larger, of the order of 2\,\kms. The error on the velocity dispersion is the random error; based on the skylines, there is an uncertainty of \around 1.7\,\kms\ in the grating resolution.  Also included is the coresponding \HI\ radial velocity and \HI\ velocity dispersion for the WiFeS field with measurements in \kms.}
\label{tab:Halpha}
\begin{tabular}{ccccclclcc} 
\hline
& & & & \Halpha\ radial &  & \Halpha\ velocity &  & \HI\ radial & \HI\ velocity \\ 
R.A. & Dec. & \Halpha\ flux  & error & velocity & error & dispersion & error & velocity & dispersion \\
\hline

  11.4350 &  -73.0839 & 1.422e-12 &   1.9e-14 &    127.4 &      0.4 &     13.4 &      0.3 &    116.5 &      8.5 \\ 
  11.6188 &  -73.1075 & 5.483e-12 &   3.8e-14 &    139.5 &      0.2 &     13.9 &      0.1 &    135.3 &     19.0 \\ 
  11.6583 &  -73.3769 & 8.214e-13 &   1.2e-14 &    143.1 &      0.4 &     11.6 &      0.3 &    132.3 &     10.8 \\ 
  11.6592 &  -73.5317 & 2.810e-12 &   1.4e-14 &    136.6 &      0.1 &     11.3 &      0.1 &    134.8 &     11.3 \\ 
  11.9275 &  -73.1242 & 2.514e-12 &   2.6e-14 &    143.1 &      0.3 &     13.9 &      0.2 &    136.1 &     19.6 \\ 
  11.9729 &  -73.1706 & 1.830e-12 &   2.2e-14 &    151.8 &      0.4 &     17.8 &      0.4 &    134.5 &     13.6 \\ 
  12.0067 &  -73.2933 & 1.481e-12 &   1.7e-14 &    147.7 &      0.2 &     11.2 &      0.2 &    135.6 &     12.9 \\ 
  12.0992 &  -73.2581 & 1.426e-12 &   1.7e-14 &    148.1 &      0.3 &     12.3 &      0.3 &    136.1 &     14.0 \\ 
  12.2171 &  -73.1303 & 7.599e-13 &   1.6e-14 &    140.1 &      0.6 &     16.0 &      0.5 &    138.0 &     14.0 \\ 
  12.4113 &  -73.4364 & 3.932e-13 &   1.3e-14 &    165.5 &      1.1 &     23.6 &      1.1 &    161.1 &     13.4 \\ 
  12.4792 &  -72.5769 & 3.865e-12 &   2.8e-14 &    172.1 &      0.2 &     16.0 &      0.2 &    156.7 &      8.5 \\ 
  12.6108 &  -72.8869 & 5.651e-12 &   3.6e-14 &    138.8 &      0.1 &     11.5 &      0.1 &    139.1 &     16.9 \\ 
  12.6800 &  -72.8411 & 1.263e-12 &   2.0e-14 &    149.9 &      0.4 &     11.9 &      0.3 &    144.9 &     17.1 \\ 
  12.6946 &  -72.7908 & 2.048e-12 &   2.5e-14 &    153.7 &      0.3 &     15.4 &      0.3 &    141.5 &     16.2 \\ 
  12.7017 &  -73.3475 & 6.526e-13 &   1.1e-14 &    164.8 &      0.4 &     13.3 &      0.4 &    153.6 &     17.5 \\ 
  12.8971 &  -72.8578 & 4.509e-13 &   1.5e-14 &    157.1 &      1.0 &     15.1 &      0.8 &    146.0 &     14.7 \\ 
  12.9800 &  -73.4706 & 7.444e-13 &   8.4e-15 &    171.5 &      0.3 &     17.0 &      0.3 &    168.4 &     12.1 \\ 
  13.0171 &  -73.2258 & 4.581e-13 &   8.9e-15 &    166.1 &      0.5 &     15.7 &      0.5 &    149.8 &     12.6 \\ 
  13.1417 &  -72.6714 & 5.633e-13 &   1.7e-14 &    163.2 &      1.1 &     24.7 &      1.1 &    164.3 &     10.7 \\ 
  13.5129 &  -72.7078 & 5.909e-13 &   2.0e-14 &    148.8 &      0.9 &     20.1 &      1.0 &    133.4 &     11.0 \\ 
  13.5758 &  -72.8061 & 9.320e-12 &   4.5e-14 &    156.8 &      0.1 &     18.1 &      0.1 &    131.5 &     14.3 \\ 
  14.0262 &  -72.3106 & 7.083e-13 &   1.9e-14 &    152.1 &      0.7 &     14.5 &      0.7 &    136.0 &      8.1 \\ 
  14.5550 &  -72.6542 & 6.963e-12 &   3.9e-14 &    175.5 &      0.2 &     20.9 &      0.1 &    177.8 &      8.7 \\ 
  14.6446 &  -72.1928 & 4.071e-12 &   3.5e-14 &    174.9 &      0.2 &     11.1 &      0.2 &    175.9 &     11.0 \\ 
  14.8387 &  -72.2317 & 2.196e-12 &   2.5e-14 &    178.2 &      0.3 &     14.1 &      0.3 &    172.2 &     11.7 \\ 
  14.8938 &  -72.1164 & 2.736e-12 &   2.4e-14 &    173.8 &      0.2 &     11.6 &      0.2 &    172.0 &      8.5 \\ 
  15.0613 &  -72.1533 & 8.796e-13 &   2.0e-14 &    176.8 &      0.6 &     12.3 &      0.6 &    173.7 &     10.1 \\ 
  15.6550 &  -72.4214 & 6.768e-13 &   1.8e-14 &    169.6 &      1.1 &     28.8 &      0.9 &    179.5 &     10.1 \\ 
  15.8417 &  -72.0603 & 5.120e-12 &   3.0e-14 &    189.2 &      0.2 &     13.1 &      0.1 &    179.8 &     10.1 \\ 
  16.0954 &  -72.1781 & 3.782e-13 &   1.8e-14 &    185.0 &      1.7 &     23.0 &      1.6 &    181.0 &     12.2 \\ 
  16.1363 &  -71.9011 & 6.471e-12 &   2.1e-14 &    187.6 &      0.1 &     15.5 &      0.1 &    179.2 &     12.5 \\ 
  16.3108 &  -72.0133 & 2.479e-12 &   2.2e-14 &    205.7 &      0.3 &     24.3 &      0.3 &    188.1 &     12.5 \\ 
  16.3925 &  -72.1453 & 6.136e-13 &   1.5e-14 &    188.2 &      0.8 &     18.0 &      0.7 &    183.2 &     13.7 \\ 
  16.4583 &  -72.4792 & 3.108e-13 &   1.7e-14 &    174.7 &      1.9 &     21.8 &      1.8 &    187.3 &     12.7 \\ 
  17.0583 &  -71.9986 & 1.416e-12 &   1.7e-14 &    182.3 &      0.3 &     10.5 &      0.3 &    186.0 &     13.4 \\ 
  17.3242 &  -73.1944 & 4.429e-13 &   1.3e-14 &    175.6 &      0.8 &     11.7 &      0.7 &    169.9 &     11.4 \\ 
  17.7554 &  -72.7114 & 2.753e-13 &   1.5e-14 &    157.3 &      1.3 &     13.1 &      1.4 &    139.8 &      8.3 \\ 
  18.7358 &  -73.3308 & 5.502e-12 &   3.5e-14 &    185.1 &      0.2 &     11.9 &      0.1 &    177.2 &     10.1 \\ 
  18.9608 &  -73.1817 & 6.224e-13 &   9.7e-15 &    186.1 &      0.4 &     12.5 &      0.4 &    180.7 &      9.7 \\ 
  21.0287 &  -73.1544 & 8.602e-14 &   9.9e-15 &    169.7 &      2.7 &     12.9 &      3.1 &    157.2 &     11.0 \\ 
  22.3675 &  -73.5628 & 4.163e-12 &   3.1e-14 &    183.9 &      0.2 &     14.1 &      0.2 &    181.6 &      8.6 \\ 

\end{tabular}
\end{table*}


\begin{table*}
\centering
\caption{The line fluxes and metallicity values of the 29 WiFeS significant fields. R.A.\ and Dec.\ are in degrees,  fluxes in erg\,s$^{-1}$\,cm$^{-2}$\,\AA\!\!$^{-1}$, and metallicities in dex.}
\label{tab:metallicity}
\begin{tabular}{cccccccccccl} 
\hline
     &      & \Hbeta &       & [OIII]$\lambda$5007  &       & \Halpha &       & [NII]$\lambda$6584  &       & O3N2        &       \\
     R.A. & Dec.      & flux      & error     & flux      & error     & flux      & error     & flux      & error     & metallicity & error  \\
\hline 
  11.4350 &  -73.0839 & 3.745e-13 &   3.8e-14 & 1.909e-12 &   5.3e-14 & 1.422e-12 &   1.9e-14 & 6.069e-14 &   1.2e-14 &    8.065 &     0.032 \\ 
  11.6188 &  -73.1075 & 1.600e-12 &   5.6e-14 & 7.582e-12 &   9.3e-14 & 5.483e-12 &   3.8e-14 & 1.289e-13 &   1.3e-14 &    7.993 &     0.015 \\ 
  11.6583 &  -73.3769 & 1.789e-13 &   2.0e-14 & 1.087e-13 &   1.3e-14 & 8.214e-13 &   1.2e-14 & 6.744e-14 &   7.6e-15 &    8.452 &     0.028 \\ 
  11.6592 &  -73.5317 & 8.437e-13 &   1.4e-14 & 2.869e-12 &   1.9e-14 & 2.810e-12 &   1.4e-14 & 8.573e-14 &   5.9e-15 &    8.075 &     0.010 \\ 
  11.9275 &  -73.1242 & 6.743e-13 &   5.4e-14 & 1.560e-12 &   5.2e-14 & 2.514e-12 &   2.6e-14 & 9.593e-14 &   1.2e-14 &    8.160 &     0.021 \\ 
  11.9729 &  -73.1706 & 5.220e-13 &   2.8e-14 & 1.325e-12 &   2.8e-14 & 1.830e-12 &   2.2e-14 & 7.876e-14 &   1.4e-14 &    8.163 &     0.026 \\ 
  12.0067 &  -73.2933 & 3.830e-13 &   2.6e-14 & 4.332e-13 &   2.0e-14 & 1.481e-12 &   1.7e-14 & 1.034e-13 &   8.4e-15 &    8.343 &     0.016 \\ 
  12.2171 &  -73.1303 & 1.817e-13 &   2.0e-14 & 3.300e-13 &   1.9e-14 & 7.599e-13 &   1.6e-14 & 5.823e-14 &   9.9e-15 &    8.290 &     0.029 \\ 
  12.4792 &  -72.5769 & 8.921e-13 &   5.8e-14 & 3.129e-13 &   4.6e-14 & 3.865e-12 &   2.8e-14 & 2.505e-13 &   1.6e-14 &    8.495 &     0.024 \\ 
  12.6108 &  -72.8869 & 1.670e-12 &   5.9e-14 & 5.192e-12 &   7.1e-14 & 5.651e-12 &   3.6e-14 & 1.269e-13 &   1.3e-14 &    8.045 &     0.015 \\ 
  12.6800 &  -72.8411 & 3.110e-13 &   9.0e-14 & 3.810e-13 &   3.7e-14 & 1.263e-12 &   2.0e-14 & 9.395e-14 &   1.3e-14 &    8.341 &     0.047 \\ 
  12.6946 &  -72.7908 & 5.414e-13 &   6.4e-14 & 2.010e-12 &   5.6e-14 & 2.048e-12 &   2.5e-14 & 5.932e-14 &   1.4e-14 &    8.056 &     0.037 \\ 
  12.7017 &  -73.3475 & 1.858e-13 &   1.2e-14 & 4.456e-13 &   1.3e-14 & 6.526e-13 &   1.1e-14 & 3.654e-14 &   6.0e-15 &    8.208 &     0.025 \\ 
  12.9800 &  -73.4706 & 1.879e-13 &   8.6e-15 & 2.771e-13 &   8.0e-15 & 7.444e-13 &   8.4e-15 & 5.300e-14 &   5.2e-15 &    8.309 &     0.016 \\ 
  13.0171 &  -73.2258 & 1.146e-13 &   1.1e-14 & 2.895e-13 &   9.1e-15 & 4.581e-13 &   8.9e-15 & 2.959e-14 &   4.8e-15 &    8.221 &     0.026 \\ 
  13.5758 &  -72.8061 & 2.039e-12 &   8.7e-14 & 4.149e-12 &   8.9e-14 & 9.320e-12 &   4.5e-14 & 6.208e-13 &   2.3e-14 &    8.255 &     0.008 \\ 
  14.5550 &  -72.6542 & 1.684e-12 &   3.1e-14 & 2.138e-12 &   3.2e-14 & 6.963e-12 &   3.9e-14 & 3.861e-13 &   1.9e-14 &    8.295 &     0.008 \\ 
  14.6446 &  -72.1928 & 1.264e-12 &   6.8e-14 & 5.353e-12 &   9.5e-14 & 4.071e-12 &   3.5e-14 & 1.118e-13 &   1.3e-14 &    8.030 &     0.018 \\ 
  14.8387 &  -72.2317 & 6.381e-13 &   5.3e-14 & 2.766e-12 &   7.1e-14 & 2.196e-12 &   2.5e-14 & 9.167e-14 &   1.4e-14 &    8.085 &     0.025 \\ 
  14.8938 &  -72.1164 & 8.905e-13 &   4.6e-14 & 5.335e-12 &   6.0e-14 & 2.736e-12 &   2.4e-14 & 4.963e-14 &   1.0e-14 &    7.924 &     0.029 \\ 
  15.6550 &  -72.4214 & 1.602e-13 &   2.3e-14 & 2.423e-13 &   2.2e-14 & 6.768e-13 &   1.8e-14 & 5.557e-14 &   1.1e-14 &    8.325 &     0.037 \\ 
  15.8417 &  -72.0603 & 1.710e-12 &   1.3e-13 & 1.173e-11 &   8.8e-14 & 5.120e-12 &   3.0e-14 & 8.679e-14 &   1.1e-14 &    7.896 &     0.020 \\ 
  16.1363 &  -71.9011 & 1.623e-12 &   1.6e-14 & 4.117e-12 &   1.9e-14 & 6.471e-12 &   2.1e-14 & 2.753e-13 &   1.0e-14 &    8.162 &     0.005 \\ 
  16.3108 &  -72.0133 & 7.202e-13 &   5.0e-14 & 1.740e-12 &   4.9e-14 & 2.479e-12 &   2.2e-14 & 9.817e-14 &   1.4e-14 &    8.159 &     0.023 \\ 
  16.3925 &  -72.1453 & 1.540e-13 &   3.5e-14 & 3.341e-13 &   2.2e-14 & 6.136e-13 &   1.5e-14 & 3.551e-14 &   9.7e-15 &    8.226 &     0.050 \\ 
  17.3242 &  -73.1944 & 1.269e-13 &   1.8e-14 & 2.917e-13 &   1.9e-14 & 4.429e-13 &   1.3e-14 & 2.722e-14 &   7.8e-15 &    8.227 &     0.045 \\ 
  18.4875 &  -73.3025 & 2.011e-12 &   6.1e-14 & 6.495e-12 &   7.7e-14 & 6.514e-12 &   3.5e-14 & 1.986e-13 &   1.4e-14 &    8.082 &     0.010 \\ 
  18.7358 &  -73.3308 & 1.717e-12 &   6.5e-14 & 5.376e-12 &   8.7e-14 & 5.502e-12 &   3.5e-14 & 2.285e-13 &   1.2e-14 &    8.129 &     0.010 \\ 
  22.3675 &  -73.5628 & 1.331e-12 &   6.1e-14 & 7.290e-12 &   8.8e-14 & 4.163e-12 &   3.1e-14 & 4.643e-14 &   1.4e-14 &    7.869 &     0.041 \\ 
\end{tabular}
\end{table*}


\begin{table*}
\centering
\caption{The line fluxes and extinctions of the 28 significant WiFeS fields. R.A.\ and Dec.\ are in degrees, fluxes in erg\,s$^{-1}$\,cm$^{-2}$\,\AA\!\!$^{-1}$, and extinctions in magnitudes.}
\label{tab:EBV}
\begin{tabular}{ccccccll} 
\hline
R.A. & Dec. & \Hbeta\ flux & error & \Halpha\ flux & error & $\rm E(B-V)_{H\beta - H\alpha}$ & error \\ 
\hline
  11.4350 &  -73.0839 & 3.745e-13 &   3.8e-14 & 1.422e-12 &   1.9e-14 &   0.218 &     0.077 \\ 
  11.6188 &  -73.1075 & 1.600e-12 &   5.6e-14 & 5.483e-12 &   3.8e-14 &   0.139 &     0.027 \\ 
  11.6583 &  -73.3769 & 1.789e-13 &   2.0e-14 & 8.214e-13 &   1.2e-14 &   0.364 &     0.085 \\ 
  11.6592 &  -73.5317 & 8.437e-13 &   1.4e-14 & 2.810e-12 &   1.4e-14 &   0.117 &     0.012 \\ 
  11.9275 &  -73.1242 & 6.743e-13 &   5.4e-14 & 2.514e-12 &   2.6e-14 &   0.204 &     0.062 \\ 
  11.9729 &  -73.1706 & 5.220e-13 &   2.8e-14 & 1.830e-12 &   2.2e-14 &   0.157 &     0.042 \\ 
  12.0067 &  -73.2933 & 3.830e-13 &   2.6e-14 & 1.481e-12 &   1.7e-14 &   0.232 &     0.052 \\ 
  12.2171 &  -73.1303 & 1.817e-13 &   2.0e-14 & 7.599e-13 &   1.6e-14 &   0.293 &     0.083 \\ 
  12.4792 &  -72.5769 & 8.921e-13 &   5.8e-14 & 3.865e-12 &   2.8e-14 &   0.320 &     0.050 \\ 
  12.6108 &  -72.8869 & 1.670e-12 &   5.9e-14 & 5.651e-12 &   3.6e-14 &   0.130 &     0.027 \\ 
  12.6946 &  -72.7908 & 5.414e-13 &   6.4e-14 & 2.048e-12 &   2.5e-14 &   0.215 &     0.091 \\ 
  12.7017 &  -73.3475 & 1.858e-13 &   1.2e-14 & 6.526e-13 &   1.1e-14 &   0.158 &     0.050 \\ 
  12.9800 &  -73.4706 & 1.879e-13 &   8.6e-15 & 7.444e-13 &   8.4e-15 &   0.251 &     0.035 \\ 
  13.0171 &  -73.2258 & 1.146e-13 &   1.1e-14 & 4.581e-13 &   8.9e-15 &   0.258 &     0.073 \\ 
  13.5758 &  -72.8061 & 2.039e-12 &   8.7e-14 & 9.320e-12 &   4.5e-14 &   0.361 &     0.033 \\ 
  14.5550 &  -72.6542 & 1.684e-12 &   3.1e-14 & 6.963e-12 &   3.9e-14 &   0.284 &     0.014 \\ 
  14.6446 &  -72.1928 & 1.264e-12 &   6.8e-14 & 4.071e-12 &   3.5e-14 &   0.091 &     0.041 \\ 
  14.8387 &  -72.2317 & 6.381e-13 &   5.3e-14 & 2.196e-12 &   2.5e-14 &   0.143 &     0.064 \\ 
  14.8938 &  -72.1164 & 8.905e-13 &   4.6e-14 & 2.736e-12 &   2.4e-14 &   0.055 &     0.040 \\ 
  15.8417 &  -72.0603 & 1.710e-12 &   1.3e-13 & 5.120e-12 &   3.0e-14 &   0.035 &     0.057 \\ 
  16.1363 &  -71.9011 & 1.623e-12 &   1.6e-14 & 6.471e-12 &   2.1e-14 &   0.256 &     0.008 \\ 
  16.3108 &  -72.0133 & 7.202e-13 &   5.0e-14 & 2.479e-12 &   2.2e-14 &   0.143 &     0.054 \\ 
  17.0583 &  -71.9986 & 4.213e-13 &   4.4e-14 & 1.416e-12 &   1.7e-14 &   0.124 &     0.081 \\ 
  18.4875 &  -73.3025 & 2.011e-12 &   6.1e-14 & 6.514e-12 &   3.5e-14 &   0.096 &     0.023 \\ 
  18.7358 &  -73.3308 & 1.717e-12 &   6.5e-14 & 5.502e-12 &   3.5e-14 &   0.088 &     0.029 \\ 
  22.3675 &  -73.5628 & 1.331e-12 &   6.1e-14 & 4.163e-12 &   3.1e-14 &   0.069 &     0.035 \\ 
\end{tabular}
\end{table*}


\bsp	
\label{lastpage}
\end{document}